\documentclass[11pt]{article}
\usepackage{amsfonts,tikz,ulem,cite
}
\usepackage{subcaption}
\usepackage{amsmath}
\usepackage{amssymb}
\usepackage{graphicx}
\usepackage{color}
\usepackage{braket}
\usepackage[utf8]{inputenc}
\usepackage{hyperref} 

\pdfoutput=1

\def \be {\begin{equation}}
\def \ee {\end{equation}}
\def \ba {\begin{aligned}}
\def \ea {\end{aligned}}
\def \bea {\begin{eqnarray}}
\def \eea {\end{eqnarray}}

\begin{document}
\begin{titlepage}
\begin{flushright}
TIT/HEP-708\\
USTC-ICTS/PCFT-25-38\\
October, 2025
\end{flushright}
\vspace{0.5cm}
\begin{center}
{\Large \bf Thermodynamic Bethe ansatz and wall crossing for deformed supersymmetric quantum mechanics}

\lineskip .75em
\vskip 2.5cm
{Katsushi Ito$^{a,}$\footnote{ito@th.phys.titech.ac.jp}
, Hongfei Shu$^{b,}$\footnote{ shu@zzu.edu.cn
} and Jingjing Yang$^{a,c,d,}$\footnote{jj150632@mail.ustc.edu.cn}
}
\vskip 2.5em
 {\normalsize\it 
 $^{a}$ Department of Physics, Institute of Science Tokyo,
Tokyo, 152-8551, Japan\\
$^{b}$School of Physics and Microelectronics,
Zhengzhou University, Zhengzhou, Henan 450001, China\\

$^{c}$Interdisciplinary Center for Theoretical Study,
University of Science and Technology of China, Hefei, Anhui 230026, China\\
$^{d}$Peng Huanwu Center for Fundamental Theory, Hefei, Anhui 230026, China
}
\vskip 3.0em
\end{center}
\begin{abstract}
We study the deformed supersymmetric quantum mechanics with a polynomial superpotential with $\hbar$ correction. 
In the minimal chamber, where all turning points are real and distinct, it was shown that the exact Wentzel--Kramers--Brillouin periods obey the ${\mathbb Z}_4$-extended  thermodynamic Bethe ansatz (TBA) equations of the undeformed potential. By changing the energy parameter above/below the critical points, the turning points become complex, and the moduli are outside of the minimal chamber. We study the wall crossing of the ${\mathbb Z}_4$-extended TBA equations by this change of moduli and show that the ${\mathbb Z}_4$ structure is preserved after the wall crossing.  In particular, the TBA equations for the cubic superpotential are studied in detail, where there are two chambers (minimal and maximal). At the maximally symmetric point in the maximal chamber, the TBA system becomes the two sets of the $D_3$-type TBA equations, which are regarded as the ${\mathbb Z}_4$ extension of the $A_3/{\mathbb Z}_2$-type TBA equation.
\end{abstract}
\end{titlepage}

\tableofcontents
\newpage

\section{Introduction}
Resurgence theory has drawn much attention recently as a non-perturbative framework \cite{ecalle}, primarily because most perturbative series in quantum field theory are asymptotic and divergent \cite{Zinn-Justin:1980oco}. Resummation techniques have been used to transform divergent series into an analytic function. Moreover, resurgence theory may enable us to extract the non-perturbative contribution by tracking the connections between different perturbative series.  The exact WKB method in one-dimensional quantum mechanics is one of the well-studied examples. In this approach, the energy spectrum can be computed by solving the exact quantization condition satisfied by the exact WKB periods \cite{voros-quartic}, whose analytic structure is determined by their asymptotics and discontinuity. 

On the other hand, the connection between the spectral problem of ordinary differential equations and functional relations in quantum integrable models is found in \cite{Dorey:1998pt,Bazhanov:1998wj}, known as the ODE/IM correspondence. 
Recently, a relation between the exact WKB periods for one-dimensional quantum mechanics and the Y-functions as the solution of the thermodynamic Bethe ansatz (TBA) equations in quantum integrable models was uncovered \cite{Ito:2018eon}.
It was shown that the TBA equations of the associated integrable model can reproduce the asymptotics and discontinuity of the WKB periods, which thus provide the exact WKB periods.

The connection between exact WKB periods and TBA equations was initially established in one-dimensional quantum mechanics with a polynomial potential. The corresponding Schr\"odinger equation can also be interpreted as the quantum Seiberg--Witten (SW) curve of $A_r$-type Argyres--Douglas (AD) theory in the Nekrasov--Shatashvili (NS) limit of the omega background \cite{Nekrasov:2009rc,Ito:2017ypt}. In this interpretation, the WKB periods correspond to central charges of BPS particles in the AD theory. The corresponding TBA equations can also be obtained by taking the conformal limit of the non-linear integral equations for the central charges of the BPS particles in the Gaiotto--Moore--Neitzke (GMN) formalism \cite{Gaiotto:2009hg,Gaiotto:2014bza}, which will be called GMN TBA in this paper.

Based on the ODE/IM correspondence and the GMN TBA, the correspondence between the exact WKB method and the TBA equations has been generalized to the quantum SW curve for ${\cal N}=2$ $SU(2)$ gauge theory \cite{Grassi:2019coc,Fioravanti:2019vxi,Grassi:2021wpw,Fioravanti:2022bqf,Hao:2025azt}, $(A_r,A_N)$-type AD theory \cite{Ito:2021sjo,Ito:2021boh}, and $D$-type AD theory \cite{Ito:2019jio,Gabai:2021dfi,Ito:2024wxw}. See also \cite{Hollands:2019wbr,Dumas:2020zoz} for other examples. Given the BPS spectrum in a certain chamber of SW theory \cite{Alim:2011ae,Maruyoshi:2013fwa,Longhi:2016rjt}, one can write down the GMN TBA equations. The conformal limit of these then yields the TBA equations governing the exact WKB periods of the quantum SW curve. See \cite{Ito:2025pfo} for review and references therein.

Recently, the resurgence theory, e.g. the exact WKB method, was applied to the deformed supersymmetric quantum mechanics\cite{Behtash:2015loa,Fujimori:2017osz,Kamata:2021jrs}. The model includes a bosonic field and some Grassmann fields with a superpotential. After integrating out the fermionic degrees of freedom, the resulting model has an effective potential with $\hbar$ correction.
In the previous work \cite{Ito:2024nlt}, we studied the relation between the exact WKB periods in the deformed supersymmetric quantum mechanics and the TBA equations. A key distinction from earlier studies is that this Schr\"odinger equation does not appear as a quantum SW curve, which suggests it may lie outside the scope of the conventional GMN formalism. Applying the exact WKB method and the ODE/IM correspondence to this model, we derived a set of $\mathbb{Z}_4$-extended TBA equations. 
A notable feature of the effective potential is the emergence of an $\hbar$ corrected term, which distinguishes it significantly from the standard exact WKB analysis for the Schr\"odinger equation with only classical potential. In particular, odd power terms will appear in the $\hbar$ expansion of the WKB periods.
In this case, the discontinuity structure of the WKB periods can be understood from the $\mathbb{Z}_4$-extended TBA equations.
For the cubic superpotential with real and distinct turning points, these TBA equations split into two independent $D_3$-type TBA equations. Notably, these $D_3$-type TBA systems also appear as the GMN TBA equations of the $D_3$-type AD theory in the minimal chamber. It is thus interesting to extend the TBA equations to the whole moduli space of the superpotential, where the TBA equations will be modified at the marginal stability wall of the WKB periods (BPS particles). This modification is called the wall crossing of the TBA equations.

In this paper, we present a detailed study on the wall crossing of the $\mathbb{Z}_4$-extended TBA equations to explore the correspondence between the exact WKB method and the quantum integrable models for the deformed supersymmetric quantum mechanics. This paper is organized as follows: In section \ref{sec:exact-wkb}, we review the exact WKB analysis for the Schr\"odinger equation with $\hbar$-deformed potential. We then present the $\mathbb{Z}_4$-extended TBA equations, and show the general procedure of the wall crossing of the $\mathbb{Z}_4$-extended TBA in section \ref{sec:TBA-wc}. In section \ref{sec:TBA-cubic}, we will apply the general procedure to the cubic superpotential. We work out in detail the monomial superpotential, where the effective potential becomes the quartic monomial potential with $\hbar$ correction.

\section{Exact WKB Analysis for the Schr\"odinger equation}\label{sec:exact-wkb}
In this section, we briefly review the exact WKB analysis for the Schr\"odinger equation for the effective potential with $\hbar$ correction:
\begin{equation}\label{eq:hbar-Sch}
    \Big(-\hbar^2\frac{d^2}{d x^2}+Q_0(x)+\hbar Q_1(x)\Big)\psi(x)=0,
\end{equation}
where $x$ is a complex variable. See \cite{kawai2005algebraic, DDP-97} for details. The potential terms $Q_0(x)$ and $Q_1(x)$ are assumed to be polynomials in $x$. 
In this paper, we study the deformed supersymmetric quantum mechanics, where they are determined by a superpotential $W(x)$ as
\begin{equation}
    Q_0(x)
    =(W^\prime(x))^2-2E,\quad Q_1(x)=2mW^{\prime\prime}(x).
\end{equation}
The parameter $m$ is a deformation parameter, where $m=\pm 1/2$ corresponds to the supersymmetric case.
Here we will focus on the polynomial-type superpotential $W(x)$:
 \begin{equation}\label{eq:superp1}
W(x)=x^{N+1}-\sum_{j=0}^{N}u_{j}x^{j}.
 \end{equation} 
 Here $m$ and $u_j$ can be any complex number in general.
We now study the WKB solution of Eq. \eqref{eq:hbar-Sch}, where we use the ansatz:
 \begin{equation}
     \psi(x)=\exp\Big(\frac{1}{\hbar}\int^x P(x^\prime)dx^\prime\Big).
 \end{equation}
Then $P(x)$ is shown to satisfy the Riccati equation:
 \begin{equation}
     \hbar\frac{d}{dx}P(x)+P(x)^{2}-Q_0(x)-\hbar Q_1(x)=0.
 \end{equation}
 Substituting the power expansion $P(x)=\sum_{n=0}^\infty\hbar^np_n(x)$ into the Riccati equation, one can determine $p_n$ recursively.

For the study of the spectral problem, it is important to consider the period integral $\oint P(x)dx$ on the WKB curve $\Sigma_{\rm WKB}$, which is defined by the algebraic relation of complex coordinates $(y,x)$:
 \begin{equation}
     y^2=Q_0(x).
 \end{equation}
 The solution of $y$ provides the leading order term $p_0$ of $P(x)$.
For the superpotential $W(x)$ in \eqref{eq:superp1}, $Q_0(x)$ is a polynomial of order $2N$. Then it has $2N$ zeros, denoted as $x_1,\cdots, x_{2N}$, which are called turning points. First, we assume that these turning points are all real, distinct, and ordered $x_1<\cdots<x_{2N}$. We denote the one-cycle going around the interval $[x_{s},x_{s+1}]$ by $\gamma_s$ ($s=1,\cdots,2N-1$). The one-cycle $\gamma_s$ with even $s$ (odd $s$) is also assumed to correspond to the forbidden (allowed) region, where the potential term $Q_0(x)$ becomes positive (negative). We will choose the branch cut and the orientation of each cycle such that the corresponding classical period, $\oint_{\gamma_s}p_0(x)dx$, is positive real (positive imaginary) for the forbidden cycle (allowed cycle). For a general WKB curve, periods can be obtained by deforming the locations of the turning points.

Given a one-cycle $\gamma$ on the curve, one defines the WKB period
\begin{equation}\label{eq:qp1}
    \Pi_{\gamma}=\oint_{\gamma} P(x)dx=\sum_{n=0}^\infty \hbar^n\Pi_{\gamma}^{(n)},\quad  \Pi_{\gamma}^{(n)}=\oint_{\gamma} p_n(x)dx.
\end{equation}
The classical period and its quantum corrections $\Pi^{(n)}_\gamma$ ($n\geq 0$) are expressed as an integral of a meromorphic differential and can be expanded as
\begin{equation}\label{eq:wkb-quantum-corrections}
    \Pi_{\gamma}^{(n)}=\sum_{k=0}^{2N-2}B_{k}^{(n)}\big(\oint_{\gamma}\frac{x^{k}}{y}dx\big),\quad k=0,\dots,2N-2,
\end{equation}
 where $B_k^{(n)}$ are constants and $\oint_{\gamma}\frac{x^{k}}{y}dx$ are the period integrals of the basis of meromorphic differentials on the WKB curve.
It is known that the $\hbar$ expansion \eqref{eq:qp1} is an asymptotic series.
It is treated as an analytic function after the Borel transform:
\begin{equation}
    \widehat{\Pi}_\gamma(\xi)=\sum_{n=0}^\infty {\xi^n\over n!}\Pi_\gamma^{(n)},
\end{equation}
where $\xi$ is a complex variable in the Borel plane.  The asymptotic series can be recovered by the Borel resummation along the direction $\varphi$, defined by the Laplace transform:
\begin{equation}\label{eq:borel-resum}
    s_{\varphi}\left(\Pi_\gamma\right)(\hbar)=\frac{1}{\hbar}\int_0^{\infty e^{i\varphi}}e^{-{\xi\over \hbar}}\widehat{\Pi}_\gamma(\xi)d\xi .
\end{equation}
When singularities of $\widehat{\Pi}_\gamma(\xi)$ exist along the direction $\varphi$, the integral is not well-defined. We circumvent this by alternatively defining the two lateral Borel resummations as:
\begin{equation}\label{eq:lateral-borel-resum}
s_{\varphi\pm}(\Pi_\gamma)(\hbar)=\lim_{\delta\to0}s_{\varphi\pm\delta}(\Pi_\gamma)(\hbar).
\end{equation}
The lateral Borel resummation shows discontinuities of the period and characterizes the analytic structure of the exact WKB periods. For a polynomial potential without $\hbar$ correction, the discontinuities are determined by the Delabaere--Dillinger--Pham (DDP) formula \cite{ddp,Iwaki:2014vad,Iwaki-2}. It turns out the DDP formula can be expressed in the form of the non-linear integral equations called the TBA equations, which have been studied in the context of BPS spectra in ${\cal N}=2$ supersymmetric gauge theories \cite{Gaiotto:2009hg,Gaiotto:2014bza}.
In the case of a polynomial potential in the minimal chamber, which includes the case where all the turning points are real and different, the TBA equations take the form of the $A_r$-type.
However, the structure of TBA equations changes when potential parameters are outside the minimal chamber. This is called the wall crossing phenomenon.

For deformed supersymmetric quantum mechanics, the TBA equations in the minimal chamber take the form of ${\mathbb Z}_4$ extension of the undeformed ones.
The main purpose of the present work is to explore the wall crossing of the ${\mathbb Z}_4$-extended TBA equations for the deformed supersymmetric quantum mechanics which includes the $\hbar$ correction in the potential.

\section{$\mathbb{Z}_4$-extended TBA and wall crossing}\label{sec:TBA-wc}
In this section, we first present the $\mathbb{Z}_4$-extended TBA equations in the minimal chamber for the deformed supersymmetric quantum mechanics with superpotential \eqref{eq:superp1} using the ODE/IM correspondence \cite{Ito:2024nlt}. We then provide a general procedure for wall crossing of these $\mathbb{Z}_4$-extended TBA to extend them to the whole space of moduli parameters. 

\subsection{$\mathbb{Z}_4$-extended TBA from the ODE/IM correspondence}

We first study the asymptotic solutions of the ODE at infinity and the Y-system constructed from the Wronskians of the solutions. At infinity, there are two kinds of asymptotic solutions: growing and decaying. Since only the decaying solution is uniquely defined, we will focus on the fastest decaying solution.
For positive real $\hbar$, the fastest decaying solution of \eqref{eq:hbar-Sch} at infinity $y(x,u_{j},E,m,\hbar)$ around the positive real axis is
\begin{equation}
    y(x,u_{j},E,m,\hbar)\sim\frac{1}{\sqrt{2i}}(\hbar^{-\frac{1}{N+1}}x)^{n_{N}}\exp\Big(-\frac{1}{\hbar}\frac{x^{N+1}}{N+1}\Big),
\end{equation}
where $n_{N}=-\frac{N}{2}-B_{N+1}$. Note that $E$ and $m$ will appear in the subleading order of $y(x,u_{j},E,m,\hbar)$, which can be obtained by solving \eqref{eq:hbar-Sch}. Here, $B_n$ is defined by
\begin{equation}
    \sqrt{Q_{0}(x)+\hbar Q_{1}(x)}\sim x^{N}\big(1+\sum_{n=1}^{\infty}B_{n}\hbar^{\frac{n}{N+1}}x^{-n}\big).
\end{equation}
The solution $ y(x,u_{j},E,m,\hbar)$ is subdominant in sector ${\cal S}_0:|{\rm arg}(x)|<\frac{\pi}{2N+2}$. The subdominant solution in sector ${\cal S}_k: |{\rm arg}(x)-\frac{2k\pi}{2N+2}|<\frac{\pi}{2N+2}$ ($k\in \mathbb{Z}$) can be obtained from $y(x,u_{j},E,m,\hbar)$ by
\begin{equation}\label{eq:yk}
    y_{k}(x,u_{j},E,m,\hbar)=\omega^{\frac{k}{2}}y(\omega^{-k}x,\omega^{-k(N+1-j)}u_{j},\omega^{-k2N}E,\omega^{-k(N+1)}m,\hbar),
\end{equation}
where $\omega=e^{\frac{2\pi i}{2N+2}}$. Here we have used the Symanzik rotation of \eqref{eq:hbar-Sch}, $(x,u_j,E,m,\hbar)\to (\omega x,\omega^{N+1-j}u_j,\omega^{2N}E,\omega^{N+1}m,\hbar)$, which will keep the Schr\"odinger equation \eqref{eq:hbar-Sch} invariant, but rotate the solution.

So far, we have fixed $\hbar$ to be positive and real. The solutions can be continued analytically to complex $\hbar$. In this case,  $y_k$ defined in \eqref{eq:yk} is the subdominant solution in ${\cal S}_k: |{\rm arg}(x)-\frac{{\rm arg}(\hbar)}{N+1}-\frac{2k\pi}{2N+2}|<\frac{\pi}{2N+2}$. Note that the subdominant solutions $y_k$ \eqref{eq:yk} can be obtained by rotation of $\hbar$ and $m$ as \cite{Ito:2024nlt}
\begin{equation}
    y_{k}(x,u_{j},E,m,\hbar)=\omega^{\frac{k}{2}}y(x,u_{j},E,e^{-i\pi k}m,e^{i\pi k}\hbar).
\end{equation}

The Wronskian of $y_{k_1}$ and $y_{k_2}$ is defined by
\begin{equation}
    W_{k_{1},k_{2}}(u_{j},E,m,\hbar):=y_{k_{1}}\partial_{x}y_{k_{2}}-y_{k_{2}}\partial_{x}y_{k_{1}}.
\end{equation}
Following the procedure in \cite{Ito:2024nlt}, we can further introduce the Y-functions:
\begin{equation}
    \begin{aligned}
       {Y}_{2N-2k}(u_{j},E,{m},\hbar)&=\frac{{W}_{-k,k}{W}_{-k-1,k+1}}{{W}_{-k-1,-k}{W}_{k,k+1}}(u_{j},E,{m},\hbar),\\
        {Y}_{2N-(2k+1)}(u_{j},E,{m},\hbar)&=\frac{{W}_{-k-1,k}{W}_{-k-2,k+1}}{{W}_{-k-2,-k-1}{W}_{k,k+1}}(u_{j},E,{m},\hbar),
    \end{aligned}
\end{equation}
which satisfy the Y-system
\begin{equation}
    {Y}_{s}(e^{\frac{\pi i}{2}}\hbar,e^{-\frac{\pi i}{2}}{m}){Y}_{s}(e^{-\frac{\pi i}{2}}\hbar,e^{\frac{\pi i}{2}}{m})=\big(1+{Y}_{s-1}\big)\big(1+{Y}_{s+1}\big)(\hbar,{m}).
\end{equation}
Here we have omitted the arguments $u_j$ and $E$ above, which are fixed in the relation. Let us introduce the $\mathbb{Z}_4$-extended Y-functions $Y_{a,s}(\hbar)$ ($a=0,1,2,3\ {\rm mod}\, 4$) to distinguish their $m$-dependence:
\begin{equation}
   Y_{a,s}(\hbar):= Y_{s}(\hbar,e^{a\frac{\pi i}{2}} m),
\end{equation}
where we have omitted the argument $m$ on the left hand side. We aim to convert the Y-system into integral equations that depend on the parameters $u_j, E$, and $m$.
We first focus on the minimal chamber. In this case, we can identify the one-cycle $\gamma_s$ on the WKB curve with the Y-function $Y_s$ \cite{Alday:2010vh}. Let us introduce the spectral parameter $\theta=-\log\hbar$. Then, from the analyticity in the strip $|{\rm Im}\theta|<{\pi\over2}$ and the asymptotic behavior $Y_s(\theta)\rightarrow -m_s e^\theta$ at $\theta\rightarrow +\infty$, the Y-system can be converted into the $\mathbb{Z}_4$-extended TBA equations
\begin{equation}
    \begin{aligned}
        \log Y_{a,s}=&-m_{s}e^{\theta}+m_{a,s}^{(\frac{1}{2})}+K_{+}\star L_{a+1,s-1}+K_{+}\star L_{a+1,s+1}\\&+K_{-}\star L_{a+3,s-1}+K_{-}\star L_{a+3,s+1},\quad a\equiv a+4,\quad s=1,\cdots,2N-1,
    \end{aligned}
\end{equation}
where $\star$ denotes the convolution: $f\star g(\theta)\equiv \int d\theta' f(\theta-\theta')g(\theta')$, $L_{a,s}=\log\big(1+Y_{a,s}\big)$ and $Y_0=Y_{2N}=0$. The kernels $K_{\pm}$ are given by
\begin{equation}
    K_\pm=\frac{1}{4\pi}\Big(\frac{1}{\cosh\theta}\pm i\frac{\sinh\theta}{\cosh\theta}\Big).
\end{equation}
The driving terms are determined by the asymptotics of the Y-functions at $\theta\rightarrow +\infty$ as 
\begin{equation}
    \begin{aligned}
        m_{2k-1}&=\frac{1}{i}\oint_{\gamma_{2k-1}}p_0 dx,\quad m_{2k}=\oint_{\gamma_{2k}}p_0 dx,\quad  
        m_{a,s}^{(\frac{1}{2})}
        =-i^a\oint_{\gamma_s}p_1 dx, 
    \end{aligned}
\end{equation}
where $1\leq 2k-1,2k\leq 2N-1$ and $\gamma_s$ are the ones defined in the previous section. We first suppose that $m_s$ is real and positive, which corresponds to certain regions of the moduli parameters of the potential $Q_0$.

Note that $m_{a,s}^{(\frac{1}{2})}$ plays the role of chemical potential in the TBA equations. 
Setting $m=0$, $Y_{a,s}$ for any $a$ are identified, which leads to the $A_{2N-1}$-type TBA equations \cite{Ito:2018eon}. Compared to the undeformed TBA in \cite{Ito:2018eon}, the kernel $K(\theta)=1/(2\pi \cosh(\theta))$ is decomposed into two kernels $K_\pm$. Due to the $\tanh(\theta)$ term, the kernels approach to nonzero constants at large $\theta$, which will cause difficulty in the numerical study. Expanding the kernels at large $\theta$, the term $\tanh(\theta)$ in the kernel will also lead to odd power terms in the $\hbar$-expansion of WKB periods, which do not appear in the Schr\"odinger equation without $\hbar$-deformed potential. Moreover, singularities will appear in the kernel of TBA equations if we shift $\theta$ by $\pm \pi i/2$, which leads to discontinuity of $\log Y(\theta\pm \pi i/2)$\footnote{This will lead to the DDP formula for the corresponding WKB periods. Very different from the original TBA, the DDP formula differs in the directions ${\rm Im}(\theta)=\pi/2$ and ${\rm Im}(\theta)=-\pi/2$.}.

When some turning points are away from the real axis, the mass terms $m_s=|m_s|e^{i\phi_s}$ can be complex-valued, and the TBA equations become
\begin{equation}\label{eq:Z4TBA-com}
    \begin{aligned}
        \log\widetilde{Y}_{a,s}=&-|m_{s}|e^{\theta}+m_{a,s}^{(\frac{1}{2})}+K_{+;s,s-1}\star\widetilde{L}_{a+1,s-1}+K_{+;s,s+1}\star\widetilde{L}_{a+1,s+1}\\&+K_{-;s,s-1}\star\widetilde{L}_{a+3,s-1}+K_{-;s,s+1}\star\widetilde{L}_{a+3,s+1},\quad a\equiv a+4,
    \end{aligned}
\end{equation}
where $\widetilde{Y}_{a,s}=Y_{a,s}(\theta-i\phi_{s})$ and $\widetilde{L}_{a,s}=\log\big(1+\widetilde{Y}_{a,s}\big)$. The kernels are modified to be
\begin{equation}
    K_{\pm;s^{\prime},s}=K_{\pm}(\theta-i\phi_{s^{\prime}}+i\phi_{s}).
\end{equation}

Our TBA equations correspond to the kink limit of the massive quantum integral model \cite{Zamolodchikov:1989cf,Klassen:1989ui}, which is characterized by the effective central charge of the underlying CFT. The corresponding effective central charge is defined by
\begin{equation}\label{eq:ceff-Z4}
    \begin{aligned}
        c_{{\rm eff}}=&\frac{6}{\pi^{2}}\sum_{a=0}^{3}\sum_{s=1}^{2N-1}\int_{-\infty}^{\infty}d\theta|m_{s}|e^{\theta}\widetilde{L}_{a,s}(\theta).
    \end{aligned}
\end{equation}

It is critical to note that poles of the kernel $K_\pm(\theta-\theta^\prime-i\phi_{s}+i\phi_{s\pm 1})$, i.e. $\theta^\prime=\theta-i\phi_{s}+i\phi_{s\pm 1}\pm i\pi/2$, will collide with the real and positive axis of the convolution as $|\phi_s-\phi_{s\pm 1}|=\pi/2$, where one has to modify the TBA equations by picking up the contribution of poles. TBA equations \eqref{eq:Z4TBA-com} are only valid in the regions $|\phi_s-\phi_{s\pm 1}|<\pi/2$, which is denoted as the minimal chamber. In the following section, we will provide the general procedure of the wall crossing for the $\mathbb{Z}_4$-extended TBA.

\subsection{wall crossing of $\mathbb{Z}_4$-extended TBA}\label{sc:wall crossing-Z4-TBA}
Without loss of generality, we consider the wall crossing of TBA equations when  $\phi_2-\phi_1$ increases below  $\pi/2$ and goes above. In this case, we should pick up the contribution of the poles of $K_{\pm;1,2}(\theta-\theta^\prime)$ and $K_{\pm;2,1}(\theta-\theta^\prime)$, where the poles are located at $\theta^\prime=\theta-i\phi_1+i\phi_2-\frac{i\pi}{2}$ and $\theta^\prime=\theta-i\phi_2+i\phi_1+\frac{i\pi}{2}$, respectively. Picking up these contributions, we obtain the TBA equations
\begin{equation}\label{eq:TBA-res}
    \begin{aligned}
               \log\widetilde{Y}_{a,1}=&-|m_{1}|e^{\theta}+m_{a,1}^{(\frac{1}{2})}+K_{+;1,2}\star\widetilde{L}_{a+1,2}+K_{-;1,2}\star\widetilde{L}_{a+3,2}+L_{a+3,2}(\theta-i\phi_{1}-\frac{i\pi}{2}),\\
               \log\widetilde{Y}_{a,2}=&-|m_{2}|e^{\theta}+m_{a,2}^{(\frac{1}{2})}+K_{+;2,1}\star\widetilde{L}_{a+1,1}+K_{+;2,3}\star\widetilde{L}_{a+1,3}\\&+K_{-;2,1}\star\widetilde{L}_{a+3,1}+K_{-;2,3}\star\widetilde{L}_{a+3,3}+L_{a+1,1}(\theta-i\phi_{2}+\frac{i\pi}{2}),\\
       \log\widetilde{Y}_{a,3}=&-|m_{3}|e^{\theta}+m_{a,3}^{(\frac{1}{2})}+K_{+;3,2}\star\widetilde{L}_{a+1,2}+K_{+;3,4}\star\widetilde{L}_{a+1,4}\\&+K_{-;3,2}\star\widetilde{L}_{a+3,2}+K_{-;3,4}\star\widetilde{L}_{a+3,4},\\
       &\cdots,
    \end{aligned}
\end{equation}
where we used the residues of the kernels:
\begin{equation}
    \begin{aligned}
        &{\rm Res}_{\theta^{\prime}=\theta\mp i(\phi_{1}-\phi_{2}+\frac{\pi}{2})}K_{\pm}(\theta-\theta^{\prime}\mp i(\phi_{1}-\phi_{2}))=0,\\
         &{\rm Res}_{\theta^{\prime}=\theta\mp i(\phi_{1}-\phi_{2}+\frac{\pi}{2})}K_{\mp}(\theta-\theta^{\prime}\mp i(\phi_{1}-\phi_{2}))=\pm\frac{i}{2\pi}.
    \end{aligned}
\end{equation}
It is useful to introduce new Y-functions
\begin{equation}
    \begin{aligned}
        Y_{a,1}^{{\rm n}}(\theta)=&\frac{Y_{a,1}(\theta)}{1+Y_{a+3,2}(\theta-\frac{i\pi}{2})},\quad Y_{a,2}^{{\rm n}}(\theta)=\frac{Y_{a,2}(\theta)}{1+Y_{a+1,1}(\theta+\frac{i\pi}{2})},\\
        Y_{a,12}^{{\rm n}}(\theta)=&\frac{Y_{a,1}(\theta)Y_{a+3,2}(\theta-\frac{i\pi}{2})}{1+Y_{a,1}(\theta)+Y_{a+3,2}(\theta-\frac{i\pi}{2})},\quad         Y_{a,s}^{{\rm n}}(\theta)=Y_{a,s}(\theta),\quad s=3,4,\cdots, 2N-1,
    \end{aligned}
\end{equation}
such that $L_{a+3,2}$, and $L_{a+1,1}$ on the right-hand side of the first two TBA equations in \eqref{eq:TBA-res} can be absorbed to the left-hand side. The definitions and the Y-system lead to
\begin{equation}
    \begin{aligned}
        &1+Y_{a,1}(\theta)=\big(1+Y_{a,1}^{{\rm n}}(\theta)\big)\big(1+Y_{a,12}^{{\rm n}}(\theta)\big),\\
        &1+Y_{a,2}(\theta)=\big(1+Y_{a,2}^{{\rm n}}(\theta)\big)\big(1+Y_{a+1,12}^{{\rm n}}(\theta+\frac{\pi i}{2})\big).
    \end{aligned}
\end{equation}
Using these relations and shifting $\theta^\prime$ in the convolutions, one obtains
\begin{equation}\label{eq:TBA-wc-gen1}
    \begin{aligned}
        \log\widetilde{Y}_{a,1}^{{\rm n}}=&-|m_{1}|e^{\theta}+m_{a,1}^{(\frac{1}{2})}\\&+K_{+;1,2}\star\widetilde{L}_{a+1,2}^{{\rm n}}+K_{+;1,12}^{[+]}\star\widetilde{L}_{a+2,12}^{{\rm n}}+K_{-;1,2}\star\widetilde{L}_{a+3,2}^{{\rm n}}+K_{-;1,12}^{[+]}\star\widetilde{L}_{a,12}^{{\rm n}},\\\log\widetilde{Y}_{a,2}^{{\rm n}}=&-|m_{2}|e^{\theta}+m_{a,2}^{(\frac{1}{2})}+K_{+;2,1}\star\widetilde{L}_{a+1,1}^{{\rm n}}+K_{+;2,12}\star\widetilde{L}_{a+1,12}^{{\rm n}}+K_{+;2,3}\star\widetilde{L}_{a+1,3}^{{\rm n}}\\&+K_{-;2,1}\star\widetilde{L}_{a+3,1}^{{\rm n}}+K_{-;2,12}\star\widetilde{L}_{a+3,12}^{{\rm n}}+K_{-;2,3}\star\widetilde{L}_{a+3,3}^{{\rm n}},\\\log\widetilde{Y}_{a,3}^{{\rm n}}=&-|m_{3}|e^{\theta}+m_{a,3}^{(\frac{1}{2})}+K_{+;3,2}\star\widetilde{L}_{a+1,2}^{{\rm n}}+K_{+;3,12}^{[+]}\star\widetilde{L}_{a+2,12}^{{\rm n}}+K_{+;3,4}\star\widetilde{L}_{a+1,4}^{{\rm n}}\\&+K_{-;3,2}\star\widetilde{L}_{a+3,2}^{{\rm n}}+K_{-;3,12}^{[+]}\star\widetilde{L}_{a,12}^{{\rm n}}+K_{-;3,4}\star\widetilde{L}_{a+3,4}^{{\rm n}},\\&\dots, 
    \end{aligned}
\end{equation}
where we defined $f^{[\pm]}=f(\theta\pm i\pi/2)$ for the kernels. To find a closed TBA system, we need the TBA equation for $Y_{a,12}^{{\rm n}}$, which can be obtained by taking the summation of 
\begin{equation}
    \begin{aligned}
        \log\frac{Y_{a,1}(\theta)}{1+Y_{a+3,2}(\theta-\frac{i\pi}{2})}=&-m_{1}e^{\theta}+m_{a,1}^{(\frac{1}{2})}+K_{+}(\theta+i\phi_{2})\star\widetilde{L}_{a+1,2}+K_{-}(\theta+i\phi_{2})\star\widetilde{L}_{a+3,2},\\
        \log\frac{Y_{a+3,2}(\theta-\frac{i\pi}{2})}{1+Y_{a,1}(\theta)}=&-m_{2}e^{\theta-\frac{i\pi}{2}}+m_{a+3,2}^{(\frac{1}{2})}+K_{+}(\theta+i\phi_{1}-\frac{i\pi}{2})\star\widetilde{L}_{a,1}+K_{+}(\theta+i\phi_{3}-\frac{i\pi}{2})\star\widetilde{L}_{a,3}\\&+K_{-}(\theta+i\phi_{1}-\frac{i\pi}{2})\star\widetilde{L}_{a+2,1}+K_{-}(\theta+i\phi_{3}-\frac{i\pi}{2})\star\widetilde{L}_{a+2,3}.
    \end{aligned}
\end{equation}
The summation of the left-hand side leads to
\begin{equation}
    \log\Big(\frac{Y_{a,1}(\theta)}{1+Y_{a+3,2}(\theta-\frac{i\pi}{2})}\frac{Y_{a+3,2}(\theta-\frac{i\pi}{2})}{1+Y_{a,1}(\theta)}\Big)=\log\frac{Y_{a,12}^{{\rm n}}(\theta)}{1+Y_{a,12}^{{\rm n}}(\theta)}.
\end{equation}
The denominator can be canceled by the terms of the summation of the right-hand side, which finally leads to
\begin{equation}\label{eq:TBA-wc-gen2}
    \begin{aligned}
       \log\widetilde{Y}_{a,12}^{{\rm n}}(\theta)=&-|m_{12}|e^{\theta}+m_{a,1}^{(\frac{1}{2})}+m_{a+3,2}^{(\frac{1}{2})}+K_{+;12,2}\star\widetilde{L}_{a+1,2}^{{\rm n}}+K_{-;12,2}\star\widetilde{L}_{a+3,2}^{{\rm n}}\\&+K_{+;12,1}^{[-]}\star\widetilde{L}_{a,1}^{{\rm n}}+K_{-;12,1}^{[-]}\star\widetilde{L}_{a+2,1}^{{\rm n}}+K_{+;12,3}^{[-]}\star\widetilde{L}_{a,3}^{{\rm n}}+K_{-;12,3}^{[-]}\star\widetilde{L}_{a+2,3}^{{\rm n}},
    \end{aligned}
\end{equation}
where $m_{12}=m_{1}-im_{2}=|m_{12}|e^{i\phi_{12}}$ and $\widetilde{L}_{a,12}^{\rm n}=\log(1+\widetilde{Y}^{\rm n}_{a,12})$. \eqref{eq:TBA-wc-gen1} and \eqref{eq:TBA-wc-gen2} provide a closed TBA system in the chamber crossing $\phi_2-\phi_1=\pi/2$. In this process, the effective central charge \eqref{eq:ceff-Z4} becomes
\begin{equation}
    \begin{aligned}
        c_{{\rm eff}}=&\frac{6}{\pi^{2}}\int_{-\infty}^{\infty}d\theta e^{\theta}\Big\{\sum_{a=0}^{3}\sum_{s=1}^{2N-1}|m_{s}|\widetilde{L}_{a,s}^{{\rm n}}(\theta)\\&+|m_{1}|\log\big(1+Y_{a,12}^{{\rm n}}(\theta-i\phi_{1})\big)+|m_{2}|\log\big(1+Y_{a+1,12}^{{\rm n}}(\theta-i\phi_{2}+\frac{\pi i}{2})\big)\Big\}\\
        =&\frac{6}{\pi^{2}}\int_{-\infty}^{\infty}d\theta e^{\theta}\Big\{\sum_{a=0}^{3}\sum_{s=1}^{2N-1}|m_{s}|\widetilde{L}_{a,s}^{{\rm n}}(\theta)+|m_{12}|\widetilde{L}_{a,12}^{{\rm n}}\Big\},
    \end{aligned}
\end{equation}
where we have shifted the argument of the integrals associated with $Y_{12}^{\rm n}$. This provides the definition of the effective central charge for TBA equations \eqref{eq:TBA-wc-gen2} in the context of quantum integrable models. It is critical to note that the value of the effective central charge is invariant under the process of wall crossing. The wall crossing of ${\mathbb Z}_4$-extended TBA associated with other poles of the kernel can be evaluated similarly.

\section{TBA for the cubic superpotential}\label{sec:TBA-cubic}

In this section, we illustrate our general procedure of wall crossing with the case of cubic superpotential, where the Schr\"odinger equation \eqref{eq:hbar-Sch} becomes
\begin{equation}\label{eq:Sch-eq-cubic}
    \Big(-\hbar^{2}\frac{d^{2}}{dx^{2}}+\big(x^{2}-u_1\big)^{2}-2E+4\hbar mx\Big)\psi(x)=0.
\end{equation}
Let us begin with the case where the turning points are real and labelled by $-a<-b<b<a$
\begin{equation}
    a=\sqrt{u_1+\sqrt{2E}},\quad b=\sqrt{u_1-\sqrt{2E}}.
\end{equation} 
The corresponding classical WKB periods are
\begin{equation}
    \begin{aligned}
        im_1=\Pi_{\gamma_1}^{(0)}=2\int_{-a}^{-b}y dx,\quad
        m_2=\Pi_{\gamma_2}^{(0)}=2\int_{b}^{-b}y dx,\quad
       im_3= \Pi_{\gamma_3}^{(0)}=2\int_{b}^{a}y dx\,,
    \end{aligned}
\end{equation}
where $m_1, m_2$ and $m_3$ are real positive. This corresponds to the TBA equations in the minimal chamber, which will be shown soon. They can be expressed as 
\begin{equation}
    \begin{split}
        &m_1=m_3=\frac{2a^3}{3}\left[\left(1+k'^2\right)\mathbb{E}(k)-2k'^2\mathbb{K}(k)\right],\\
        &m_2=\frac{4a^3}{3}\left[\left(1+k'^2\right)\mathbb{E}(k')-k^2\mathbb{K}(k')\right],
    \end{split}
\end{equation}
where $ k^2=1-\frac{b^2}{a^2}$ and $(k')^2=1-k^2$. $\mathbb{K}(k)$ and $\mathbb{E}(k)$ are complete elliptic integrals of the first and second kinds, respectively. The quantum corrections to the classical periods can be evaluated via the Picard--Fuchs technique as shown in \eqref{eq:wkb-quantum-corrections}. For the potential in \eqref{eq:Sch-eq-cubic}, the coefficients $B_k^{(n)}$ depend only on the parameters $u_1$, $E$, and $m$. They can be determined systematically, with the first few orders given by
\begin{equation}
    \begin{split}
        &B_0^{(0)}=\frac{2}{3} \left(u_1^2-2 E\right), \quad B_1^{(0)}=0, \quad B_2^{(0)}=-\frac{2 u_1}{3},\\
        &B_0^{(1)}=0, \quad B_1^{(1)}=2m, \quad B_2^{(1)}=0,\\
        &B_0^{(2)}=\frac{\left(12 m^2-1\right) u_1}{24 E}, \quad B_1^{(2)}=0, \quad B_2^{(2)}=\frac{1}{24} \left(\frac{1-12 m^2}{E}+\frac{1}{2 E-u_1^2}\right).
    \end{split}
\end{equation}
These coefficients remain unchanged under wall crossing. We further find that all coefficients $B_k^{(2n+1)}$ vanish for $n>0$, i.e., $\Pi_\gamma^{(2n+1)}=0$. This is a distinctive property of the cubic superpotential, which is closely related to the simplification of the corresponding $\mathbb{Z}_4$-extended TBA equations, as discussed below.

\begin{figure}
  \begin{subfigure}{0.4\textwidth}
\includegraphics[width=\linewidth]{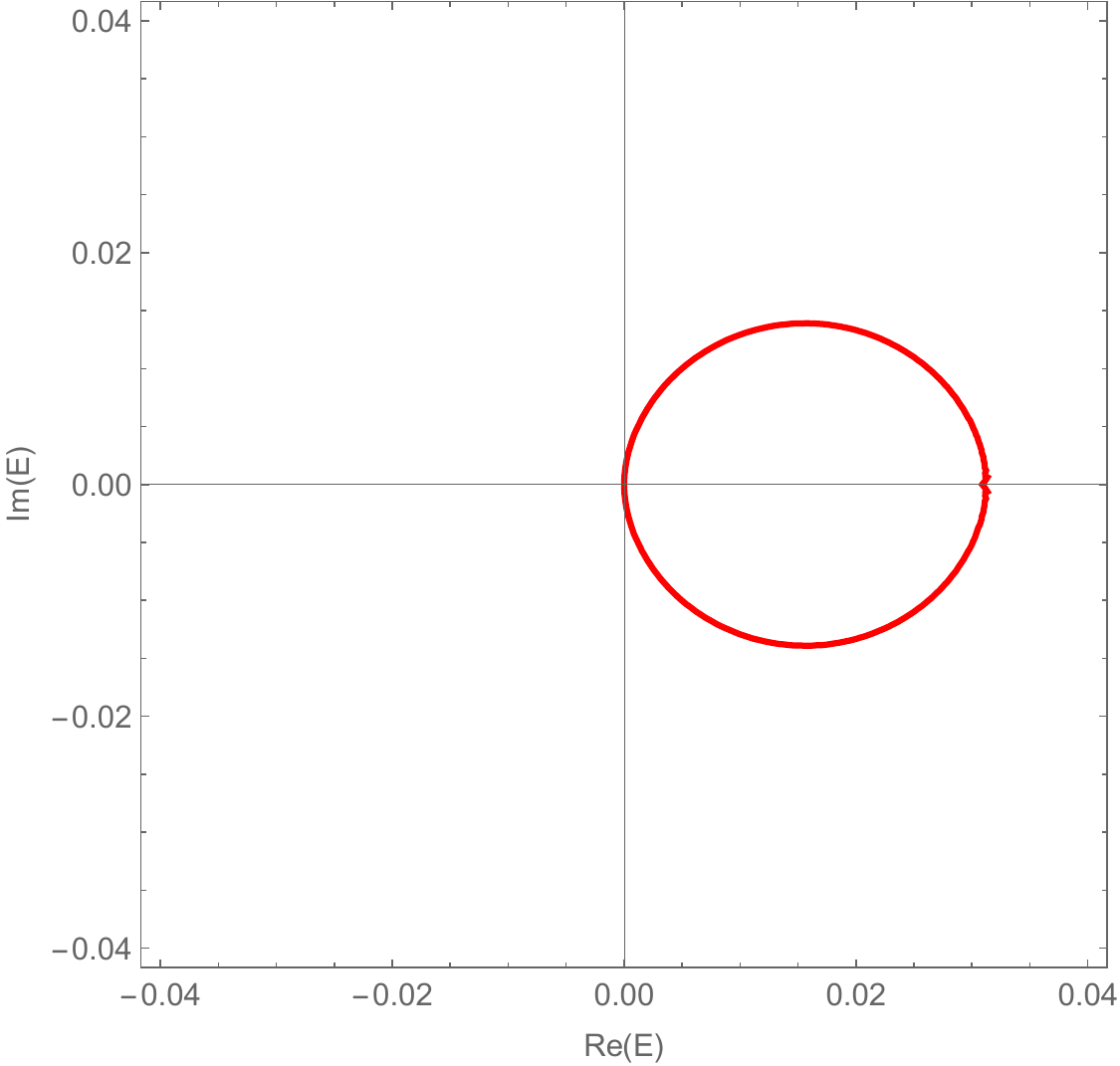}
    \caption{}
  \end{subfigure}
  \quad
    \begin{subfigure}{0.385\textwidth}
 \includegraphics[width=\linewidth]{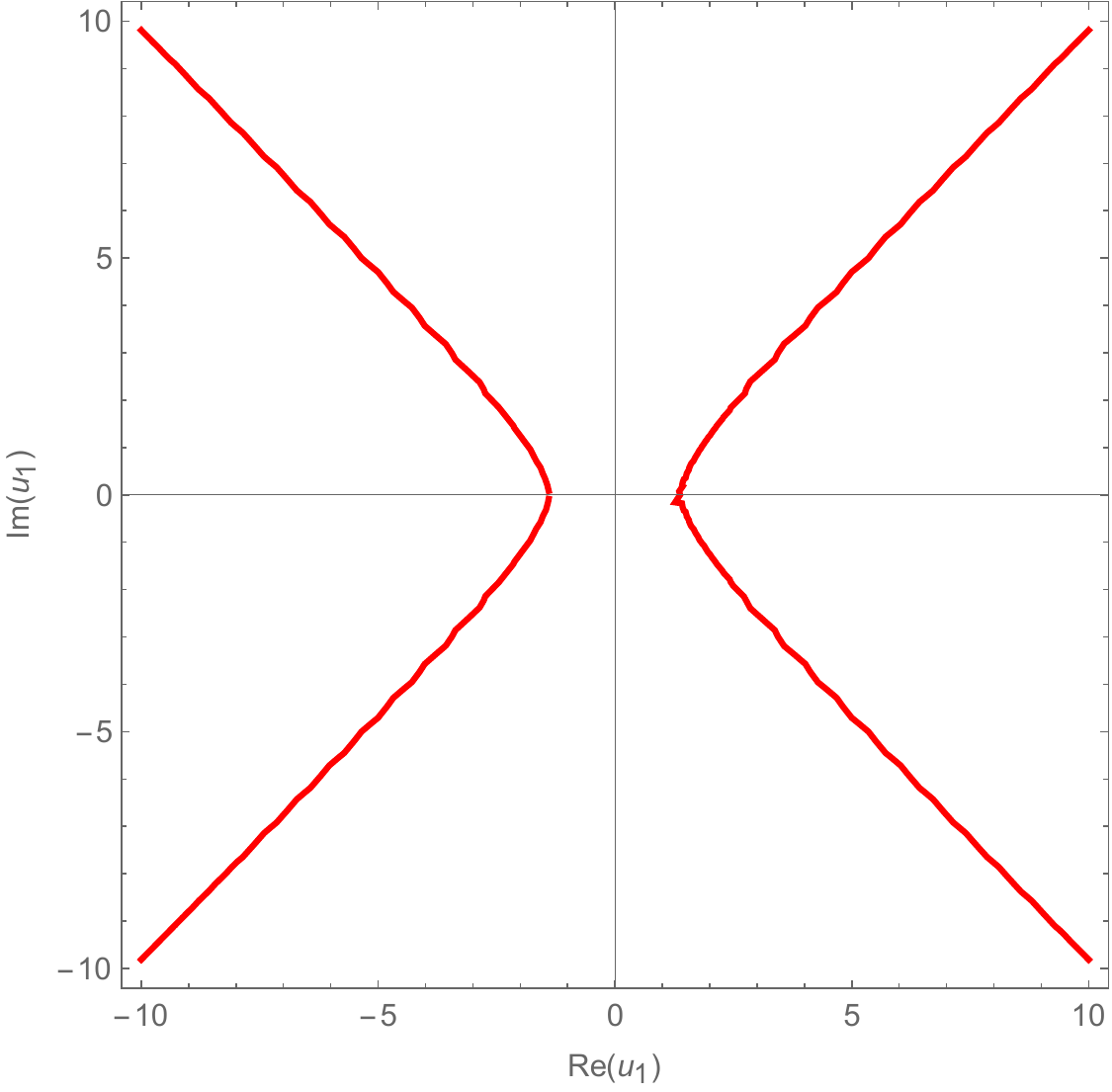}
    \caption{}     
    \end{subfigure}
    \caption{In Fig.(a),  the wall on the $E$ moduli space for a fixed value of $u_1= 1/4$ is shown. The region inside the red curve denotes the minimal chamber, while the region outside corresponds to the maximal chamber. The wall on the $u_1$ moduli space for the fixed value $E=31/32$ is shown in Fig.(b). Here the minimal chamber extends from the red line to infinity, while the maximal chamber is the bounded region near the origin.  }
    \label{fig:wall}
\end{figure}

The marginal stability wall on the moduli parameter space $(u_1,E)$ is located at
\begin{equation}
    {\rm Im}(\Pi_{\gamma_1}^{(0)}/\Pi_{\gamma_2}^{(0)})=0.
\end{equation}
In Fig. \ref{fig:wall}, the walls on the $E$ and $u_1$ moduli spaces are shown, respectively. 
In both cases, only the minimal chamber and the maximal chamber appear due to the parity symmetry $x\to -x$ in $Q_0(x)$.

\begin{figure}
\tikzset{every picture/.style={line width=0.75pt}} 

\begin{tikzpicture}[x=0.75pt,y=0.75pt,yscale=-1,xscale=1]

\draw   (29.83,82.54) .. controls (29.83,74.45) and (48.39,67.9) .. (71.28,67.9) .. controls (94.17,67.9) and (112.73,74.45) .. (112.73,82.54) .. controls (112.73,90.62) and (94.17,97.17) .. (71.28,97.17) .. controls (48.39,97.17) and (29.83,90.62) .. (29.83,82.54) -- cycle ;

\draw   (147.2,83.31) .. controls (147.2,75.22) and (165.75,68.67) .. (188.65,68.67) .. controls (211.54,68.67) and (230.1,75.22) .. (230.1,83.31) .. controls (230.1,91.39) and (211.54,97.94) .. (188.65,97.94) .. controls (165.75,97.94) and (147.2,91.39) .. (147.2,83.31) -- cycle ;

\draw   (424.01,34.18) .. controls (430.12,40.25) and (416.36,58.05) .. (393.28,73.94) .. controls (370.2,89.83) and (346.54,97.78) .. (340.43,91.71) .. controls (334.32,85.64) and (348.07,67.84) .. (371.16,51.95) .. controls (394.24,36.06) and (417.9,28.1) .. (424.01,34.18) -- cycle ;

\draw   (450.86,84.08) .. controls (450.86,75.99) and (469.42,69.44) .. (492.31,69.44) .. controls (515.21,69.44) and (533.76,75.99) .. (533.76,84.08) .. controls (533.76,92.16) and (515.21,98.71) .. (492.31,98.71) .. controls (469.42,98.71) and (450.86,92.16) .. (450.86,84.08) -- cycle ;

\draw   (455.92,266.2) .. controls (450.52,259.91) and (463.37,244.69) .. (484.62,232.18) .. controls (505.87,219.68) and (527.48,214.64) .. (532.88,220.92) .. controls (538.29,227.2) and (525.44,242.43) .. (504.19,254.93) .. controls (482.94,267.44) and (461.33,272.48) .. (455.92,266.2) -- cycle ;

\draw   (344.09,235.71) .. controls (338.22,230.08) and (349.8,213.74) .. (369.95,199.23) .. controls (390.11,184.72) and (411.2,177.52) .. (417.07,183.16) .. controls (422.93,188.79) and (411.35,205.13) .. (391.2,219.64) .. controls (371.05,234.15) and (349.95,241.35) .. (344.09,235.71) -- cycle ;

\draw   (34.13,238.4) .. controls (29.73,233.84) and (47.37,216.17) .. (73.52,198.93) .. controls (99.67,181.69) and (124.43,171.4) .. (128.83,175.96) .. controls (133.22,180.52) and (115.58,198.19) .. (89.43,215.43) .. controls (63.28,232.68) and (38.52,242.96) .. (34.13,238.4) -- cycle ;

\draw   (112.49,267.98) .. controls (109.47,262.73) and (132.07,248.63) .. (162.97,236.49) .. controls (193.87,224.35) and (221.37,218.76) .. (224.38,224.02) .. controls (227.4,229.27) and (204.8,243.37) .. (173.9,255.51) .. controls (143,267.65) and (115.5,273.23) .. (112.49,267.98) -- cycle ;

\draw    (412.83,125.17) -- (412.83,173.09) ;
\draw [shift={(412.83,175.09)}, rotate = 270] [color={rgb, 255:red, 0; green, 0; blue, 0 }  ][line width=0.75]    (10.93,-3.29) .. controls (6.95,-1.4) and (3.31,-0.3) .. (0,0) .. controls (3.31,0.3) and (6.95,1.4) .. (10.93,3.29)   ;

\draw    (252.63,91.84) -- (312.83,91.84) ;
\draw [shift={(314.83,91.84)}, rotate = 180] [color={rgb, 255:red, 0; green, 0; blue, 0 }  ][line width=0.75]    (10.93,-3.29) .. controls (6.95,-1.4) and (3.31,-0.3) .. (0,0) .. controls (3.31,0.3) and (6.95,1.4) .. (10.93,3.29)   ;

\draw    (314.03,241.09) -- (253.83,241.09) ;
\draw [shift={(251.83,241.09)}, rotate = 360] [color={rgb, 255:red, 0; green, 0; blue, 0 }  ][line width=0.75]    (10.93,-3.29) .. controls (6.95,-1.4) and (3.31,-0.3) .. (0,0) .. controls (3.31,0.3) and (6.95,1.4) .. (10.93,3.29)   ;

\draw  [color={rgb, 255:red, 208; green, 2; blue, 27 }  ,draw opacity=1 ][fill={rgb, 255:red, 208; green, 2; blue, 27 }  ,fill opacity=1 ] (99.65,84.54) .. controls (99.65,83.54) and (100.55,82.72) .. (101.65,82.72) .. controls (102.75,82.72) and (103.65,83.54) .. (103.65,84.54) .. controls (103.65,85.55) and (102.75,86.37) .. (101.65,86.37) .. controls (100.55,86.37) and (99.65,85.55) .. (99.65,84.54) -- cycle ;

\draw  [color={rgb, 255:red, 208; green, 2; blue, 27 }  ,draw opacity=1 ][fill={rgb, 255:red, 208; green, 2; blue, 27 }  ,fill opacity=1 ] (158.65,84.54) .. controls (158.65,83.54) and (159.55,82.72) .. (160.65,82.72) .. controls (161.75,82.72) and (162.65,83.54) .. (162.65,84.54) .. controls (162.65,85.55) and (161.75,86.37) .. (160.65,86.37) .. controls (159.55,86.37) and (158.65,85.55) .. (158.65,84.54) -- cycle ;

\draw  [color={rgb, 255:red, 208; green, 2; blue, 27 }  ,draw opacity=1 ][fill={rgb, 255:red, 208; green, 2; blue, 27 }  ,fill opacity=1 ] (215.65,85.54) .. controls (215.65,84.54) and (216.55,83.72) .. (217.65,83.72) .. controls (218.75,83.72) and (219.65,84.54) .. (219.65,85.54) .. controls (219.65,86.55) and (218.75,87.37) .. (217.65,87.37) .. controls (216.55,87.37) and (215.65,86.55) .. (215.65,85.54) -- cycle ;

\draw  [color={rgb, 255:red, 208; green, 2; blue, 27 }  ,draw opacity=1 ][fill={rgb, 255:red, 208; green, 2; blue, 27 }  ,fill opacity=1 ] (353.65,85.54) .. controls (353.65,84.54) and (354.55,83.72) .. (355.65,83.72) .. controls (356.75,83.72) and (357.65,84.54) .. (357.65,85.54) .. controls (357.65,86.55) and (356.75,87.37) .. (355.65,87.37) .. controls (354.55,87.37) and (353.65,86.55) .. (353.65,85.54) -- cycle ;

\draw  [color={rgb, 255:red, 208; green, 2; blue, 27 }  ,draw opacity=1 ][fill={rgb, 255:red, 208; green, 2; blue, 27 }  ,fill opacity=1 ] (408.65,46.54) .. controls (408.65,45.54) and (409.55,44.72) .. (410.65,44.72) .. controls (411.75,44.72) and (412.65,45.54) .. (412.65,46.54) .. controls (412.65,47.55) and (411.75,48.37) .. (410.65,48.37) .. controls (409.55,48.37) and (408.65,47.55) .. (408.65,46.54) -- cycle ;

\draw  [color={rgb, 255:red, 208; green, 2; blue, 27 }  ,draw opacity=1 ][fill={rgb, 255:red, 208; green, 2; blue, 27 }  ,fill opacity=1 ] (464.65,86.54) .. controls (464.65,85.54) and (465.55,84.72) .. (466.65,84.72) .. controls (467.75,84.72) and (468.65,85.54) .. (468.65,86.54) .. controls (468.65,87.55) and (467.75,88.37) .. (466.65,88.37) .. controls (465.55,88.37) and (464.65,87.55) .. (464.65,86.54) -- cycle ;

\draw  [color={rgb, 255:red, 208; green, 2; blue, 27 }  ,draw opacity=1 ][fill={rgb, 255:red, 208; green, 2; blue, 27 }  ,fill opacity=1 ] (507.65,86.54) .. controls (507.65,85.54) and (508.55,84.72) .. (509.65,84.72) .. controls (510.75,84.72) and (511.65,85.54) .. (511.65,86.54) .. controls (511.65,87.55) and (510.75,88.37) .. (509.65,88.37) .. controls (508.55,88.37) and (507.65,87.55) .. (507.65,86.54) -- cycle ;

\draw  [color={rgb, 255:red, 208; green, 2; blue, 27 }  ,draw opacity=1 ][fill={rgb, 255:red, 208; green, 2; blue, 27 }  ,fill opacity=1 ] (404.65,192.54) .. controls (404.65,191.54) and (405.55,190.72) .. (406.65,190.72) .. controls (407.75,190.72) and (408.65,191.54) .. (408.65,192.54) .. controls (408.65,193.55) and (407.75,194.37) .. (406.65,194.37) .. controls (405.55,194.37) and (404.65,193.55) .. (404.65,192.54) -- cycle ;

\draw  [color={rgb, 255:red, 208; green, 2; blue, 27 }  ,draw opacity=1 ][fill={rgb, 255:red, 208; green, 2; blue, 27 }  ,fill opacity=1 ] (350.65,230.54) .. controls (350.65,229.54) and (351.55,228.72) .. (352.65,228.72) .. controls (353.75,228.72) and (354.65,229.54) .. (354.65,230.54) .. controls (354.65,231.55) and (353.75,232.37) .. (352.65,232.37) .. controls (351.55,232.37) and (350.65,231.55) .. (350.65,230.54) -- cycle ;

\draw  [color={rgb, 255:red, 208; green, 2; blue, 27 }  ,draw opacity=1 ][fill={rgb, 255:red, 208; green, 2; blue, 27 }  ,fill opacity=1 ] (463.65,259.54) .. controls (463.65,258.54) and (464.55,257.72) .. (465.65,257.72) .. controls (466.75,257.72) and (467.65,258.54) .. (467.65,259.54) .. controls (467.65,260.55) and (466.75,261.37) .. (465.65,261.37) .. controls (464.55,261.37) and (463.65,260.55) .. (463.65,259.54) -- cycle ;

\draw  [color={rgb, 255:red, 208; green, 2; blue, 27 }  ,draw opacity=1 ][fill={rgb, 255:red, 208; green, 2; blue, 27 }  ,fill opacity=1 ] (516.65,226.54) .. controls (516.65,225.54) and (517.55,224.72) .. (518.65,224.72) .. controls (519.75,224.72) and (520.65,225.54) .. (520.65,226.54) .. controls (520.65,227.55) and (519.75,228.37) .. (518.65,228.37) .. controls (517.55,228.37) and (516.65,227.55) .. (516.65,226.54) -- cycle ;

\draw  [color={rgb, 255:red, 208; green, 2; blue, 27 }  ,draw opacity=1 ][fill={rgb, 255:red, 208; green, 2; blue, 27 }  ,fill opacity=1 ] (117.65,184.54) .. controls (117.65,183.54) and (118.55,182.72) .. (119.65,182.72) .. controls (120.75,182.72) and (121.65,183.54) .. (121.65,184.54) .. controls (121.65,185.55) and (120.75,186.37) .. (119.65,186.37) .. controls (118.55,186.37) and (117.65,185.55) .. (117.65,184.54) -- cycle ;

\draw  [color={rgb, 255:red, 208; green, 2; blue, 27 }  ,draw opacity=1 ][fill={rgb, 255:red, 208; green, 2; blue, 27 }  ,fill opacity=1 ] (119.65,264.54) .. controls (119.65,263.54) and (120.55,262.72) .. (121.65,262.72) .. controls (122.75,262.72) and (123.65,263.54) .. (123.65,264.54) .. controls (123.65,265.55) and (122.75,266.37) .. (121.65,266.37) .. controls (120.55,266.37) and (119.65,265.55) .. (119.65,264.54) -- cycle ;

\draw  [color={rgb, 255:red, 208; green, 2; blue, 27 }  ,draw opacity=1 ][fill={rgb, 255:red, 208; green, 2; blue, 27 }  ,fill opacity=1 ] (208.65,230.54) .. controls (208.65,229.54) and (209.55,228.72) .. (210.65,228.72) .. controls (211.75,228.72) and (212.65,229.54) .. (212.65,230.54) .. controls (212.65,231.55) and (211.75,232.37) .. (210.65,232.37) .. controls (209.55,232.37) and (208.65,231.55) .. (208.65,230.54) -- cycle ;

\draw  [color={rgb, 255:red, 208; green, 2; blue, 27 }  ,draw opacity=1 ][fill={rgb, 255:red, 208; green, 2; blue, 27 }  ,fill opacity=1 ] (43.65,230.54) .. controls (43.65,229.54) and (44.55,228.72) .. (45.65,228.72) .. controls (46.75,228.72) and (47.65,229.54) .. (47.65,230.54) .. controls (47.65,231.55) and (46.75,232.37) .. (45.65,232.37) .. controls (44.55,232.37) and (43.65,231.55) .. (43.65,230.54) -- cycle ;

\draw  [color={rgb, 255:red, 208; green, 2; blue, 27 }  ,draw opacity=1 ][fill={rgb, 255:red, 208; green, 2; blue, 27 }  ,fill opacity=1 ] (40.65,84.54) .. controls (40.65,83.54) and (41.55,82.72) .. (42.65,82.72) .. controls (43.75,82.72) and (44.65,83.54) .. (44.65,84.54) .. controls (44.65,85.55) and (43.75,86.37) .. (42.65,86.37) .. controls (41.55,86.37) and (40.65,85.55) .. (40.65,84.54) -- cycle ;

\draw [color={rgb, 255:red, 44; green, 26; blue, 182 }  ,draw opacity=1 ]   (42.65,84.54) .. controls (44.43,82.89) and (46.15,82.91) .. (47.82,84.58) .. controls (49.51,86.25) and (51.15,86.24) .. (52.74,84.55) .. controls (54.33,82.85) and (55.98,82.81) .. (57.71,84.42) .. controls (59.42,86.04) and (61.08,86) .. (62.7,84.31) .. controls (64.42,82.62) and (66.16,82.6) .. (67.92,84.26) .. controls (69.54,85.92) and (71.15,85.92) .. (72.76,84.26) .. controls (74.57,82.6) and (76.32,82.61) .. (78.02,84.28) .. controls (79.55,85.96) and (81.16,85.97) .. (82.84,84.32) .. controls (84.65,82.67) and (86.32,82.69) .. (87.84,84.37) .. controls (89.44,86.06) and (91.11,86.08) .. (92.84,84.43) .. controls (94.67,82.79) and (96.28,82.81) .. (97.68,84.5) -- (100.65,84.54) ;

\draw [color={rgb, 255:red, 44; green, 26; blue, 182 }  ,draw opacity=1 ]   (162.65,84.54) .. controls (164.41,82.89) and (166.08,82.91) .. (167.67,84.58) .. controls (169.42,86.24) and (171.09,86.22) .. (172.68,84.53) .. controls (174.28,82.8) and (175.94,82.74) .. (177.66,84.34) .. controls (179.45,85.97) and (181.15,85.95) .. (182.74,84.26) .. controls (184.44,82.59) and (186.1,82.59) .. (187.73,84.26) .. controls (189.37,85.93) and (191.13,85.94) .. (193.01,84.28) .. controls (194.51,82.63) and (196.06,82.64) .. (197.66,84.32) .. controls (199.4,86.01) and (201.23,86.03) .. (203.16,84.39) .. controls (204.81,82.74) and (206.5,82.76) .. (208.21,84.45) .. controls (209.56,86.14) and (211.2,86.16) .. (213.15,84.52) -- (214.65,84.54) ;

\draw [color={rgb, 255:red, 44; green, 26; blue, 182 }  ,draw opacity=1 ]   (355.65,87.37) .. controls (356.32,85.02) and (357.74,84.13) .. (359.9,84.72) .. controls (362.25,85.13) and (363.66,84.12) .. (364.14,81.69) .. controls (364.31,79.44) and (365.58,78.42) .. (367.93,78.64) .. controls (370.29,78.8) and (371.53,77.71) .. (371.65,75.36) .. controls (371.73,73.03) and (373.01,71.85) .. (375.49,71.83) .. controls (377.71,72.06) and (378.9,70.98) .. (379.05,68.6) .. controls (379.3,66.17) and (380.59,65.07) .. (382.92,65.29) .. controls (385.27,65.55) and (386.54,64.55) .. (386.73,62.3) .. controls (387.26,59.86) and (388.7,58.84) .. (391.05,59.24) .. controls (393.22,59.81) and (394.63,58.9) .. (395.29,56.51) .. controls (395.87,54.22) and (397.34,53.37) .. (399.69,53.94) .. controls (401.8,54.68) and (403.16,53.95) .. (403.76,51.75) .. controls (404.47,49.52) and (405.97,48.77) .. (408.25,49.51) -- (410.65,48.37) ;

\draw [color={rgb, 255:red, 44; green, 26; blue, 182 }  ,draw opacity=1 ]   (466.65,88.37) .. controls (468.35,86.71) and (470.05,86.7) .. (471.75,88.35) .. controls (473.57,89.9) and (475.11,89.5) .. (476.36,87.14) .. controls (477.93,85.22) and (479.64,85.04) .. (481.5,86.59) .. controls (483.17,88.2) and (484.87,88.16) .. (486.62,86.45) .. controls (488.19,84.77) and (489.87,84.76) .. (491.68,86.41) .. controls (493.35,88.08) and (494.92,88.09) .. (496.38,86.42) .. controls (498.38,84.77) and (500.23,84.78) .. (501.94,86.47) .. controls (503.29,88.15) and (504.94,88.17) .. (506.89,86.53) -- (507.65,86.54) ;

\draw [color={rgb, 255:red, 44; green, 26; blue, 182 }  ,draw opacity=1 ]   (352.65,230.54) .. controls (353.12,228.21) and (354.63,227.17) .. (357.16,227.44) .. controls (359.37,227.92) and (360.68,226.98) .. (361.1,224.62) .. controls (361.39,222.33) and (362.73,221.33) .. (365.12,221.6) .. controls (367.43,221.91) and (368.67,220.94) .. (368.82,218.69) .. controls (369.15,216.26) and (370.42,215.21) .. (372.63,215.54) .. controls (375.04,215.68) and (376.33,214.58) .. (376.5,212.25) .. controls (376.77,209.86) and (378.03,208.81) .. (380.3,209.12) .. controls (382.66,209.41) and (384,208.42) .. (384.32,206.14) .. controls (384.84,203.79) and (386.25,202.87) .. (388.56,203.4) .. controls (390.93,203.95) and (392.36,203.13) .. (392.87,200.96) .. controls (393.76,198.61) and (395.26,197.84) .. (397.37,198.65) .. controls (399.78,199.34) and (401.29,198.62) .. (401.89,196.51) .. controls (402.89,194.22) and (404.48,193.51) .. (406.65,194.37) -- (406.65,194.37) ;

\draw [color={rgb, 255:red, 44; green, 26; blue, 182 }  ,draw opacity=1 ]   (465.65,257.72) .. controls (466.14,255.38) and (467.64,254.36) .. (470.17,254.67) .. controls (472.34,255.21) and (473.66,254.32) .. (474.13,251.99) .. controls (474.49,249.73) and (475.84,248.81) .. (478.18,249.24) .. controls (480.59,249.63) and (481.99,248.69) .. (482.4,246.4) .. controls (482.85,244.09) and (484.24,243.17) .. (486.56,243.63) .. controls (488.75,244.19) and (490.11,243.31) .. (490.64,240.99) .. controls (491.16,238.71) and (492.61,237.83) .. (494.98,238.35) .. controls (497.26,238.96) and (498.71,238.13) .. (499.32,235.88) .. controls (499.95,233.63) and (501.38,232.84) .. (503.6,233.53) .. controls (505.95,234.16) and (507.51,233.32) .. (508.26,231.03) .. controls (508.94,228.77) and (510.35,228.01) .. (512.5,228.76) -- (516.65,226.54) ;

\draw [color={rgb, 255:red, 44; green, 26; blue, 182 }  ,draw opacity=1 ]   (45.65,232.37) .. controls (46.14,230.02) and (47.56,229.05) .. (49.9,229.46) .. controls (52.22,229.87) and (53.56,228.92) .. (53.92,226.61) .. controls (54.49,224.15) and (55.88,223.14) .. (58.11,223.57) .. controls (60.33,224) and (61.66,223.01) .. (62.09,220.6) .. controls (62.48,218.22) and (63.87,217.17) .. (66.24,217.46) .. controls (68.37,217.92) and (69.59,216.98) .. (69.89,214.65) .. controls (70.39,212.16) and (71.79,211.08) .. (74.09,211.41) .. controls (76.37,211.76) and (77.64,210.8) .. (77.89,208.51) .. controls (78.36,206.07) and (79.72,205.05) .. (81.98,205.44) .. controls (84.23,205.86) and (85.6,204.86) .. (86.11,202.45) .. controls (86.45,200.18) and (87.75,199.28) .. (90,199.75) .. controls (92.47,200.12) and (93.92,199.16) .. (94.36,196.89) .. controls (94.87,194.6) and (96.28,193.74) .. (98.57,194.3) .. controls (100.86,194.89) and (102.34,194.05) .. (103.01,191.76) .. controls (103.5,189.59) and (104.94,188.83) .. (107.31,189.46) .. controls (109.68,190.13) and (111.2,189.37) .. (111.87,187.19) .. controls (112.62,185) and (114.09,184.32) .. (116.28,185.15) -- (117.65,184.54) ;

\draw [color={rgb, 255:red, 44; green, 26; blue, 182 }  ,draw opacity=1 ]   (121.65,266.37) .. controls (122.2,264.01) and (123.66,263.07) .. (126.03,263.55) .. controls (128.33,264.11) and (129.79,263.25) .. (130.4,260.97) .. controls (131.07,258.7) and (132.53,257.92) .. (134.78,258.63) .. controls (136.97,259.39) and (138.42,258.67) .. (139.15,256.48) .. controls (139.93,254.29) and (141.54,253.57) .. (143.97,254.33) .. controls (146.06,255.26) and (147.53,254.67) .. (148.36,252.54) .. controls (149.53,250.31) and (151.14,249.7) .. (153.19,250.73) .. controls (155.2,251.79) and (156.81,251.23) .. (158.03,249.06) .. controls (158.98,246.99) and (160.45,246.52) .. (162.44,247.63) .. controls (164.69,248.68) and (166.32,248.19) .. (167.31,246.15) .. controls (168.61,244.02) and (170.24,243.54) .. (172.19,244.71) .. controls (174.13,245.88) and (175.76,245.42) .. (177.09,243.31) .. controls (178.12,241.28) and (179.75,240.81) .. (182,241.9) .. controls (183.96,243.07) and (185.46,242.63) .. (186.49,240.59) .. controls (187.82,238.46) and (189.47,237.96) .. (191.46,239.1) .. controls (193.47,240.22) and (194.98,239.74) .. (195.99,237.67) .. controls (197.28,235.49) and (198.95,234.93) .. (201,235.99) .. controls (203.09,237.02) and (204.61,236.47) .. (205.58,234.34) .. controls (206.51,232.21) and (208.05,231.61) .. (210.19,232.56) -- (210.65,232.37) ;

\draw    (92.65,83.31) .. controls (100.61,64.57) and (164.61,65.57) .. (172.83,82.48) ;

\draw  [dash pattern={on 4.5pt off 4.5pt}]  (93.65,83.31) .. controls (94.61,100.57) and (159.61,103.57) .. (171.81,85.18) ;

\draw    (398.65,52.31) .. controls (347.61,-6.22) and (531.61,86.78) .. (476.61,89.43) ;

\draw  [dash pattern={on 4.5pt off 4.5pt}]  (398.65,52.31) .. controls (426.61,75.43) and (473.61,97.43) .. (476.61,89.43) ;

\draw [color={rgb, 255:red, 208; green, 2; blue, 27 }  ,draw opacity=1 ]   (370.61,76.43) .. controls (252.61,101.26) and (519.61,105.43) .. (485.61,85.43) ;

\draw [color={rgb, 255:red, 208; green, 2; blue, 27 }  ,draw opacity=1 ] [dash pattern={on 4.5pt off 4.5pt}]  (370.61,76.43) .. controls (400.61,73.26) and (460.61,70.26) .. (485.61,85.43) ;

\draw [color={rgb, 255:red, 208; green, 2; blue, 27 }  ,draw opacity=1 ]   (399.61,194.65) .. controls (354.57,153) and (604.61,230.43) .. (510.61,234) ;

\draw [color={rgb, 255:red, 208; green, 2; blue, 27 }  ,draw opacity=1 ] [dash pattern={on 4.5pt off 4.5pt}]  (405.61,199.65) .. controls (455.61,222.87) and (442.61,216.65) .. (498.61,231.37) ;

\draw    (393.65,200.31) .. controls (367.61,152.43) and (467.61,225.43) .. (476.61,248.43) ;

\draw  [dash pattern={on 4.5pt off 4.5pt}]  (394.65,203.31) .. controls (416.61,233.43) and (502.61,303.43) .. (476.61,248.43) ;

\draw    (360.61,222.04) .. controls (267.61,227.65) and (515.61,293.65) .. (473.61,256.78) ;

\draw  [dash pattern={on 4.5pt off 4.5pt}]  (367.61,223.04) .. controls (397.61,228.04) and (431.61,237.65) .. (468.61,253.65) ;

\draw    (106.61,183.65) .. controls (80.61,141.65) and (294.61,248.65) .. (203.61,237.65) ;

\draw  [dash pattern={on 4.5pt off 4.5pt}]  (111.61,189.65) .. controls (139.57,212.78) and (173.61,225.65) .. (197.61,235.65) ;

\draw    (105.65,189.31) .. controls (131.61,130.65) and (146.61,230.65) .. (136.61,258.65) ;

\draw  [dash pattern={on 4.5pt off 4.5pt}]  (104.61,192.65) .. controls (98.61,236.65) and (104.61,300.65) .. (136.61,258.65) ;

\draw    (59.61,222.65) .. controls (-43.39,208.65) and (177.61,302.65) .. (136.61,258.65) ;

\draw  [dash pattern={on 4.5pt off 4.5pt}]  (59.61,222.65) .. controls (108.61,239.65) and (89.61,230.65) .. (136.61,258.65) ;

\draw [color={rgb, 255:red, 208; green, 2; blue, 27 }  ,draw opacity=1 ]   (59.61,222.65) .. controls (86.61,213.65) and (313.61,217.65) .. (195.61,238.65) ;

\draw [color={rgb, 255:red, 208; green, 2; blue, 27 }  ,draw opacity=1 ] [dash pattern={on 4.5pt off 4.5pt}]  (52.61,224.65) .. controls (-34.39,246.65) and (169.61,239.65) .. (189.61,238.65) ;

\draw (72.94,10) node [anchor=north west][inner sep=0.75pt]   [align=left] {minimal chamber};

\draw (376.91,10) node [anchor=north west][inner sep=0.75pt]   [align=left] {chamber 2};

\draw (60,177) node [anchor=north west][inner sep=0.75pt]  [rotate=-359.8] [align=left] {$\gamma_1$};

\draw (125,105) node [anchor=north west][inner sep=0.75pt]   [align=left] {$\gamma_2$};

\draw (185,105) node [anchor=north west][inner sep=0.75pt]   [align=left] {$\gamma_3$};

\draw (420,280) node [anchor=north west][inner sep=0.75pt]   [align=left] {chamber 3};

\draw (355,35) node [anchor=north west][inner sep=0.75pt]   [align=left] {$\gamma_1$};

\draw (445,35) node [anchor=north west][inner sep=0.75pt]  [rotate=-0.83] [align=left] {$\gamma_2$};

\draw (490,105) node [anchor=north west][inner sep=0.75pt]   [align=left] {$\gamma_3$};

\draw (410,105) node [anchor=north west][inner sep=0.75pt]   [align=left] {$\gamma_{12}$};

\draw (70.15,280) node [anchor=north west][inner sep=0.75pt]   [align=left] {maximal chamber};

\draw (65,105) node [anchor=north west][inner sep=0.75pt]   [align=left] {$\gamma_1$};

\draw (112,202) node [anchor=north west][inner sep=0.75pt]   [align=left] {$\gamma_2$};

\draw (140,222) node [anchor=north west][inner sep=0.75pt]   [align=left] {$\gamma_{312}$};

\draw (165,260) node [anchor=north west][inner sep=0.75pt]  [rotate=-2.28] [align=left] {$\gamma_3$};

\draw (165,177) node [anchor=north west][inner sep=0.75pt]  [rotate=-1.12] [align=left] {$\gamma_{32}$};

\draw (458.27,177) node [anchor=north west][inner sep=0.75pt]  [rotate=-358.82] [align=left] {$\gamma_{32}$};

\draw (355,177) node [anchor=north west][inner sep=0.75pt]  [rotate=-359.99] [align=left] {$\gamma_1$};

\draw (380,260) node [anchor=north west][inner sep=0.75pt]  [rotate=-0.48] [align=left] {$\gamma_{12}$};

\draw (435,215) node [anchor=north west][inner sep=0.75pt]  [rotate=-52.83] [align=left] {$\gamma_2$};

\draw (500,260) node [anchor=north west][inner sep=0.75pt]  [rotate=-359.37] [align=left] {$\gamma_3$};

\draw (380,141) node [anchor=north west][inner sep=0.75pt]   [align=left] {2nd};

\draw (265,70) node [anchor=north west][inner sep=0.75pt]   [align=left] {1st};

\draw (265,220) node [anchor=north west][inner sep=0.75pt]   [align=left] {3rd};

\draw (60,260) node [anchor=north west][inner sep=0.75pt]  [rotate=-0.41] [align=left] {$\gamma_{12}$};

\draw (240,100) node [anchor=north west][inner sep=0.75pt]   [align=left] {wall crossing};

\draw (238.8,250.08) node [anchor=north west][inner sep=0.75pt]   [align=left] {wall crossing};

\draw (415,141) node [anchor=north west][inner sep=0.75pt]   [align=left] {wall crossing};

\end{tikzpicture}
    \caption{A path from the minimal to the maximal chamber by deforming the turning points of $Q_0(x)$ (red dots). The wavy lines denote the branch cut on the WKB curve. In each chamber, the one-cycles correspond to the Y-functions. In each process of wall crossing, a new one-cycle (red cycle) and a new Y-function appear.
    }
    \label{fig:wc-path}
\end{figure}
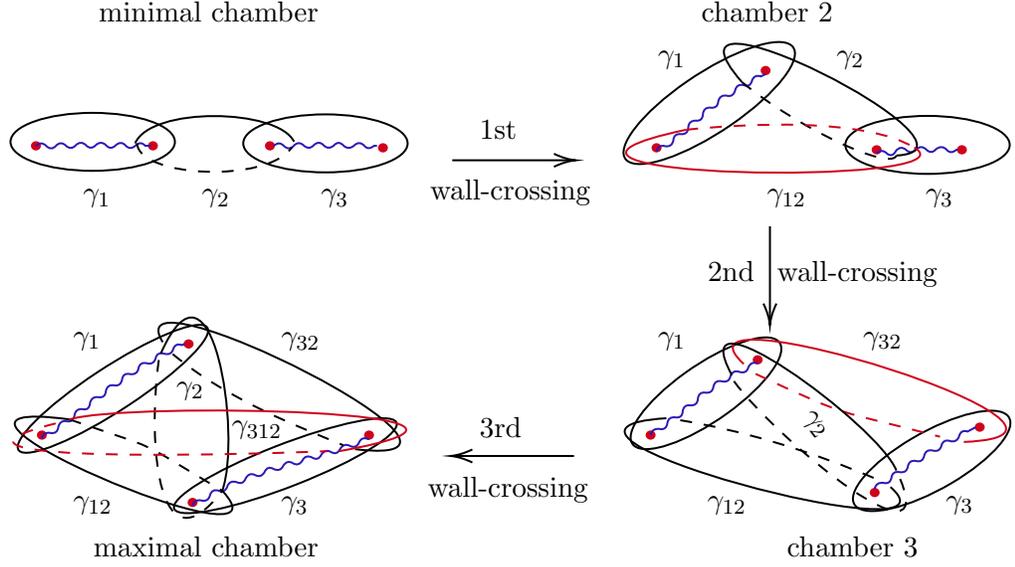

To study the wall crossing, we will break the parity symmetry of the potential $Q_0(x)$, which can be done by adding odd terms in $x$ to $Q_0(x)$. We will follow the path presented in Fig. \ref{fig:wc-path} from the minimal chamber to the maximal chamber. The first chamber is the region of the moduli space of the potential corresponding to $|\phi_2-\phi_1|<\frac{\pi}{2}$ and $|\phi_2-\phi_3|<\frac{\pi}{2}$. The second chamber is determined by $\frac{\pi}{2}<\phi_2-\phi_1<\frac{3\pi}{2}$,  $|\phi_2-\phi_3|<\frac{\pi}{2}$ and $\phi_{12}-\phi_3<0$. The third one is $\frac{\pi}{2}<\phi_2-\phi_1<\frac{3\pi}{2}$,  $\frac{\pi}{2}<\phi_2-\phi_3<\frac{3\pi}{2}$ and $\phi_{12}-\phi_3<0$. The maximal chamber is determined by $\frac{\pi}{2}<\phi_2-\phi_1<\frac{3\pi}{2}$,  $\frac{\pi}{2}<\phi_2-\phi_3<\frac{3\pi}{2}$ and $\phi_{12}-\phi_3>0$. In each chamber, a new one-cycle will be added, which corresponds to a new Y-function in the TBA system. A similar process can be found in \cite{Ito:2018eon, Ito:2024nlt}.

To obtain the TBA system for the cubic superpotential, i.e., the double well $W^{\prime 2}$, we will impose the parity symmetry on the potential. This is only possible in the minimal and maximal chambers.

\subsection{TBA in the minimal chamber}
Let us begin from the TBA equations in the minimal chamber
\begin{equation}
    \begin{aligned}
        \log\widetilde{Y}_{a,1}=&-|m_{1}|e^{\theta}+m_{a,1}^{(\frac{1}{2})}+K_{+;1,2}\star\widetilde{L}_{a+1,2}+K_{-;1,2}\star\widetilde{L}_{a+3,2},\\\log\widetilde{Y}_{a,2}=&-|m_{2}|e^{\theta}+m_{a,2}^{(\frac{1}{2})}+K_{+;2,1}\star\widetilde{L}_{a+1,1}+K_{+;2,3}\star\widetilde{L}_{a+1,3}\\&+K_{-;2,1}\star\widetilde{L}_{a+3,1}+K_{-;2,3}\star\widetilde{L}_{a+3,3},\\\log\widetilde{Y}_{a,3}=&-|m_{3}|e^{\theta}+m_{a,3}^{(\frac{1}{2})}+K_{+;3,2}\star\widetilde{L}_{a+1,2}+K_{-;3,2}\star\widetilde{L}_{a+3,2},\quad a\equiv a+4,
    \end{aligned}
\end{equation}
whose effective central charge is
\begin{equation}\label{eq:ceff-min-gen}
    c_{{\rm eff}}=\frac{6}{\pi^{2}}\sum_{a=0}^{3}\sum_{s=1}^{3}\int_{-\infty}^{\infty}d\theta|m_{s}|e^{\theta}\widetilde{L}_{a,s}(\theta).
\end{equation}
Here $m_s=|m_s|e^{i\phi_{s}}$, $\widetilde{Y}_{a,s}=Y_{a,s}(\theta-i\phi_{s})$ and $\widetilde{L}_{a,s}=\log\big(1+\widetilde{Y}_{a,s})$. Imposing the parity symmetry of the cubic superpotential, one finds
\begin{equation}
    m_1=m_3,\quad m_{a,1}^{(\frac{1}{2})}=m_{a,3}^{(\frac{1}{2})}=2\pi im,\quad m_{a,2}^{(\frac{1}{2})}=0,
\end{equation}
which leads to the identification of the Y-functions
\begin{equation}\label{eq:ident_y}
   \widetilde{Y}_{a,1}=\widetilde{Y}_{a+2,3},\quad
        \widetilde{Y}_{a,2}=\widetilde{Y}_{a+2,2}.
\end{equation}
We thus obtain a more compact form of TBA
\begin{equation}
    \begin{aligned}
        \log\widetilde{Y}_{a,1}=&-|m_{1}|e^{\theta}+2\pi im+K_{1,2}\star\widetilde{L}_{a+1,2},\\
        \log\widetilde{Y}_{a+2,1}=&-|m_{1}|e^{\theta}-2\pi im+K_{1,2}\star\widetilde{L}_{a+1,2},\\\log\widetilde{Y}_{a+1,2}=&-|m_{2}|e^{\theta}+K_{2,1}\star\Big[\widetilde{L}_{a,1}+\widetilde{L}_{a+2,1}\Big],\quad a=0,1,
    \end{aligned}
\end{equation}
where the kernel $K_{s',s}(\theta)$ is defined by $K_{s',s}(\theta)=K_{+;s',s}+K_{-;s',s}=\frac{1}{2\pi }\frac{1}{\cosh(\theta-i\phi_{s'}+i\phi_s)}$. Note that the first two TBA equations are almost the same up to the chemical potential $\pm 2\pi im$. It is thus natural to introduce the notation
\begin{equation}
    \widehat{{Y}}_{a,1}=e^{-2\pi i m}\widetilde
    {Y}_{a,1}=e^{2\pi i m}\widetilde{Y}_{a+2,1},\quad \widehat{{Y}}_{a+1,2}=\widetilde{Y}_{a+1,2},\quad  a=0,1,
\end{equation}
such that the TBA system becomes
\begin{equation}\label{eq:TBA-min}
    \begin{aligned}
        \log\widehat{{Y}}_{a,1}=-|m_{1}|e^{\theta}+K_{1,2}\star \widehat{L}_{a+1,2},\quad 
        \log\widehat{{Y}}_{a+1,2}=-|m_{2}|e^{\theta}+K_{2,1}\star  \widehat{L}_{a,1},\quad a=0,1,
    \end{aligned}
\end{equation}
where $\widehat{L}_{a,s}$ is
\begin{equation}
    \widehat{L}_{a,1}=\log\big(1+e^{2\pi im}\widehat{{Y}}_{a,1}\big)+\log\big(1+e^{-2\pi im}\widehat{{Y}}_{a,1}\big),\quad \widehat{L}_{a+1,2}=\log\big(1+\widehat{
        {Y}}_{a+1,2}\big).
\end{equation}
We thus obtain two independent $D_3$-type TBA systems in \cite{Ito:2019jio,Ito:2024nlt}. These $D_3$-type TBA systems also appear as the GMN TBA systems of $(A_1,D_3)$-type AD theory in the minimal chamber \cite{Ito:2024wxw}. As observed in \cite{Ito:2024nlt}, the TBA systems \eqref{eq:TBA-min} are only valid in the region $|m|\leq 1/2$. Beyond this region, one has to pick up the contribution of the pole at $1+\widehat{Y}=0$. For simplicity, we will focus only on the region $|m|\leq 1/2$.

Moreover, using the discontinuity formula of the kernel $K(\theta)$, the TBA system \eqref{eq:TBA-min} can be converted to the Y-system
\begin{equation}\label{eq:Y-D3}
    \begin{aligned}
    &\widehat{Y}_{a,1}(\theta-\frac{\pi i}{2})\widehat{Y}_{a,1}(\theta+\frac{\pi i}{2})=1+\widehat{Y}_{a+1,2}(\theta),\\
        &\widehat{Y}_{a+1,2}(\theta+\frac{\pi i}{2})\widehat{Y}_{a+1,2}(\theta-\frac{\pi i}{2})=\Big(1+e^{2\pi im}\widehat{Y}_{a,1}(\theta)\Big)\Big(1+e^{-2\pi im}\widehat{Y}_{a,1}(\theta)\Big),\quad a=0,1,
    \end{aligned}
\end{equation}
which are two independent $D_3$-type Y-systems. 

The effective central charge of the total system is decomposed into
\begin{equation}
    c_{\rm eff}=\widehat{c}_{{\rm eff},0}+\widehat{c}_{{\rm eff},1};
\end{equation}
where $\widehat{c}_{{\rm eff},a}$, $a=0,1$, is defined by
\begin{equation}\label{eq:ceff-min}
   \widehat{c}_{{\rm eff},a}(m):=\frac{12}{\pi^{2}}\int d\theta e^{\theta}\Big(|m_{1}|\widehat{L}_{a,1}+|m_{2}|\widehat{L}_{a+1,2}\Big).
\end{equation}

To compute this effective central charge \eqref{eq:ceff-min-gen} $\widehat{c}_{{\rm eff},a}$ analytically, we first extract the asymptotic behavior of the Y-functions for $\theta\to -\infty$, where they approach constant values $\widehat{Y}^\ast$
\begin{equation}
    \widehat{Y}_{0,1}^{\ast}=2\cos(\frac{2\pi m}{3}),\quad \widehat{Y}_{1,2}^{\ast}=\frac{\sin(2\pi m)}{\sin(\frac{2\pi m}{3})},
\end{equation}
which provides a constant solution to the Y-system \eqref{eq:Y-D3}.
Using this boundary condition, we find the effective central charge $\widehat{c}_{{\rm eff},0}(m)$
\begin{equation}
    \widehat{c}_{{\rm eff},0}(m)=\frac{12}{\pi^{2}}\Big({\cal L}_{1}\big(\frac{1}{1+\widehat{Y}_{1,2}^{\ast-1}}\big)+{\cal L}_{e^{2\pi im}}\big(\frac{e^{-2\pi im}}{1+\widehat{Y}_{0,1}^{\ast-1}}\big)+{\cal L}_{e^{-2\pi im}}\big(\frac{e^{2\pi im}}{1+\widehat{Y}_{2,1}^{\ast-1}}\big)\Big)=4(1-8m^{2}),
\end{equation}
where ${\cal L}_c(x)$ is the Rogers dilogarithm function: 
\begin{equation}
    \begin{aligned}
        {\cal L}_{c}(x)=-\frac{1}{2}\int_{0}^{x}dy\Big(\frac{c\log y}{1-cy}+\frac{\log(1-cy)}{y}\Big)=\frac{1}{2}\big(\log(x)\log(1-cx)+2{\rm Li}_{2}(cx)\big).
    \end{aligned}
\end{equation}
Similarly, one can compute the constant values for $a=1$, which lead to the effective central charge
\begin{equation}
    \widehat{c}_{{\rm eff},1}(m)=4(1+8m^{2}).
\end{equation}
The total effective central charge is thus
\begin{equation}\label{eq:total-ceff}
    c_{\rm eff}=8.
\end{equation}

We then compare the asymptotic behavior of the Y-functions with the WKB periods. As $\theta \to \infty$, the Y-functions can be expanded as:

\begin{equation}
-\log\widehat{Y}_{a,s}(\theta)\sim|m_{s}|e^{\theta}+\sum_{n=1}^{\infty}\widehat{m}_{a,s}^{(n)}e^{(1-2n)\theta}.
\end{equation}
Here, the coefficients $\widehat{m}_{a,1}^{(n)}$ and $\widehat{m}_{a+1,2}^{(n)}$ are given by:

\begin{equation}\label{eq:mass1}
\begin{aligned}
\widehat{m}_{a,1}^{(n)}=\frac{(-1)^{n}}{\pi}\int_{-\infty}^{\infty}e^{(2n-1)(\theta+i\phi_1-i\phi_2)}\widehat{L}_{a+1,2}d\theta, \quad \widehat{m}_{a+1,2}^{(n)}=\frac{(-1)^{n}}{\pi}\int_{-\infty}^{\infty}e^{(2n-1)(\theta+i\phi_2-i\phi_1)}\widehat{L}_{a,1}d\theta.
\end{aligned}
\end{equation}

These asymptotic expansions are compared with the quantum corrections to the WKB periods, defined in \eqref{eq:qp1}. We obtain
\begin{equation}\label{eq:mass_wkb}
\widehat{m}_{0,1}^{(n)}=e^{i(2n-1)(\phi_1+\frac{\pi}{2})}\Pi_{\gamma_1}^{(2n)}, \qquad \widehat{m}_{1,2}^{(n)}=e^{i(2n-1)\phi_2}\Pi_{\gamma_2}^{(2n)}.
\end{equation}
We can numerically solve the TBA equations \eqref{eq:TBA-min} with $a=0$, and verify the above identifications, as shown in Table \ref{tab:tba_wkb-min}. For real masses, where $\phi_1=\phi_2=0$, these relationships were established in \cite{Ito:2024nlt}.

It is interesting to note that the effective central charge \eqref{eq:ceff-min} can be expressed by using the expansion \eqref{eq:mass1} and the WKB periods \eqref{eq:mass_wkb}:
\begin{equation}
    \widehat{c}_{{\rm eff},a}=-\frac{12}{\pi}\big(m_{1}\widehat{m}_{a+1,2}^{(1)}e^{-i\phi_{2}}+m_{2}\widehat{m}_{a,1}^{(1)}e^{-i\phi_{1}}\big)=\frac{12i}{\pi}(\Pi_{\gamma_1}^{(0)}\Pi_{\gamma_2}^{(2)}-\Pi_{\gamma_2}^{(0)}\Pi_{\gamma_1}^{(2)}),
\end{equation}
which imposes a strong constraint on the classical part and the second-order correction of the WKB periods. This type of constraint is known as the perturbative/non-perturbative relation of the WKB periods \cite{Matone:1995rx,Codesido:2017dns,Basar:2017hpr}.

A more comprehensive identification can be established by analyzing the discontinuity structure of both the Y-functions and the Borel-resummed WKB periods. The agreement of these discontinuities, combined with the asymptotic behavior, leads to the following formulas:
\begin{equation}
-\log \Big(e^{2\pi i m}\widehat{Y}_{0,1}\big(\theta+i\phi_1+\frac{i\pi}{2}\pm i0\big)\Big)=\frac{1}{\hbar}s_{0\pm}\left(\Pi_{\gamma_1}\right)(\hbar), \quad -\log \widehat{Y}_{1,2}(\theta+i\phi_2)=\frac{1}{\hbar}s_0\left(\Pi_{\gamma_2}\right)(\hbar),
\end{equation}
where the Borel resummations are defined as in \eqref{eq:borel-resum} and \eqref{eq:lateral-borel-resum}. Alternatively, these relations can be expressed in terms of $Y_{0,1}$ and $Y_{1,2}$ as:
\begin{equation}\label{eq:TBA-WKB-relation}
-\log Y_{0,1}\left(\theta+\frac{i\pi}{2}\pm i0\right)=\frac{1}{\hbar}s_{0\pm}\left(\Pi_{\gamma_1}\right)(\hbar), \quad -\log Y_{1,2}(\theta)=\frac{1}{\hbar}s_0\left(\Pi_{\gamma_2}\right)(\hbar).
\end{equation}
The other set of TBA equations, $a=1$, in \eqref{eq:TBA-min} can be obtained from the ones of $a=0$ by the analytic continuation of the deformation parameter $m\to im$. We can find the relations between the WKB periods and the Y-functions for $a=1$:
\begin{equation}
-\log Y_{1,1}\left(\theta+\frac{i\pi}{2}\pm i0\right)=\left.\frac{1}{\hbar}s_{0\pm}\left(\Pi_{\gamma_1}\right)(\hbar)\right|_{m\to im}, \quad -\log Y_{0,2}(\theta)=\left.\frac{1}{\hbar}s_0\left(\Pi_{\gamma_2}\right)(\hbar)\right|_{m\to im},
\end{equation}
where we used \eqref{eq:ident_y}: $\widehat{Y}_{a,2}=\widehat{Y}_{a+2,2}$. The expansions of both sides lead to
\begin{equation}
\widehat{m}_{1,1}^{(n)}=\left.e^{i(2n-1)(\phi_1+\frac{\pi}{2})}\Pi_{\gamma_1}^{(2n)}\right|_{m\to im}, \quad \widehat{m}_{0,2}^{(n)}=\left.e^{i(2n-1)\phi_2}\Pi_{\gamma_2}^{(2n)}\right|_{m\to im},
\end{equation}
which have been tested numerically, similarly to Table \ref{tab:tba_wkb-min}.

\begin{table}[]
    \centering
    \begin{tabular}{|c|c|c|}\hline
    $n$&    $\widehat{m}^{(n)}_{0,1}$ & $e^{i(2n-1)(\phi_1+\pi/2)}\Pi_{\gamma_{1}}^{(2n)}$\\\hline
   $1$ & $ -0.566356132934705 + 0.824251768834188 i $ & $-0.566356132934311 + 0.824251768834206 i$ \\\hline
   $2$&$-63.545928924630 -18.895873594992 i $ & $
   -63.545928924625 - 18.895873594966 i$  \\\hline\hline
    $n$ & $\widehat{m}_{1,2}^{(n)}$ & $e^{i(2n-1)\phi_2}\Pi_{\gamma_{2}}^{(2n)}$\\\hline
     $1$ & $ 3.029657720424650 + 3.757337183886880 i $ & $3.029657720424889 + 3.757337183887550 i$ \\\hline
   $2$&$951.232049818090 - 436.302918571272 i $ & $951.232049818474 - 436.302918571300 i$ \\\hline
    \end{tabular}    \caption{
 Comparison of the coefficients in the expansions based on the TBA equations and the WKB periods at $m=2/5$ with $u_1=1/4+i/16$ and $E=1/64$. The TBA equations are solved numerically by using the discretized Fourier transformation with $2^{16}$ points and the cutoff  $(-50,50)$.}
    \label{tab:tba_wkb-min}
\end{table}

\subsection{TBA in the maximal chamber}

After the third wall crossing in Fig.\ref{fig:wc-path}, we arrive at the maximal chamber, where $4\times 6$ TBA equations are introduced---see \eqref{eq:TBA-max-gen} in Appendix \ref{sec:appA}. Compared with those in the minimal chamber, three cycles $\gamma_{12}$, $\gamma_{32}$ and $\gamma_{312}$ have been added, which correspond to three types of new Y-functions $Y_{a,12}, Y_{a,32}$ and $Y_{a,312}$, respectively, whose masses are given by 
\begin{equation}\label{eq:new-mass-max}
    m_{12}=m_{1}-im_{2}=|m_{12}|e^{i\phi_{12}},\quad m_{32}=m_{3}-im_{2}=|m_{32}|e^{i\phi_{32}},\quad m_{312}=m_{3}+m_{12}=|m_{312}|e^{i\phi_{312}}.
\end{equation}
For the cubic superpotential \eqref{eq:Sch-eq-cubic} in the maximal chamber, the parity symmetry $x\rightarrow -x$ of the potential $Q_0(x)$ implies
\begin{equation}\label{eq:mass-rel-max}
    m_1=m_3,\quad m_{12}=m_{32},\quad m_{a,1}^{(\frac{1}{2})}=m_{a,3}^{(\frac{1}{2})}=2\pi im,\quad m_2^{(\frac{1}{2})}=0,
\end{equation}
and also the identifications
\begin{equation}\label{eq:iden-max}
   \widetilde{Y}_{a,1}=\widetilde{Y}_{a+2,3},\quad\widetilde{Y}_{a,12}=\widetilde{Y}_{a+2,32},\quad \widetilde{Y}_{a,2}=\widetilde{Y}_{a+2,2},\quad\widetilde{Y}_{a,312}=\widetilde{Y}_{a+2,312}.
\end{equation}
The TBA equations \eqref{eq:TBA-max-gen} thus reduce to two independent TBA systems \eqref{eq:TBA-max-dou-1}. As in the case of the minimal chamber, we introduce 
\begin{equation}\label{eq:new-Y-max}
\begin{split}
    &\widehat{
    {Y}}_{a,1}^{(3)}=e^{-2\pi im}\widetilde{Y}_{a,1}^{(3)}=e^{2\pi im}\widetilde{Y}_{a+2,1}^{(3)},\quad\widehat{
    {Y}}_{a,12}^{(3)}=e^{-2\pi im}\widetilde{Y}_{a,12}^{(3)}=e^{2\pi im}\widetilde{Y}_{a+2,12}^{(3)},\\
    &\widehat{Y}_{a+1,2}^{(3)}=\widetilde{Y}_{a+1,2}^{(3)}, \quad \widehat{Y}_{a,312}^{(3)}=\widetilde{Y}_{a,312}^{(3)}, \quad a=0,1,
\end{split}
\end{equation}
such that the TBA systems \eqref{eq:TBA-max-dou-1} simplify to
\begin{equation}\label{eq:TBA-cubic-max}
    \begin{aligned}
        \log\widehat{Y}_{a,1}^{(3)}=&-|m_{1}|e^{\theta}+K_{1,2}\star\widehat{L}_{a+1,2}^{(3)}+K_{1,12}^{[+]}\star\widehat{L}_{a,12}^{(3)}+K_{1,312}^{[+]}\star\widehat{L}_{a,312}^{(3)},\\\log\widehat{Y}_{a,12}^{(3)}=&-|m_{12}|e^{\theta}+K_{12,1}^{[-]}\star\widehat{L}_{a,1}^{(3)}+K_{12,2}\star\widehat{L}_{a+1,2}^{(3)}+K_{12,312}^{[-]}\star\widehat{L}_{a,312}^{(3)},\\\log\widehat{Y}_{a+1,2}^{(3)}=&-|m_{2}|e^{\theta}+K_{2,1}\star\widehat{L}_{a,1}^{(3)}+K_{2,12}\star\widehat{L}_{a,12}^{(3)}+2K_{2,312}\star\widehat{L}_{a,312}^{(3)},\\\log\widehat{Y}_{a,312}^{(3)}=&-|m_{312}|e^{\theta}+K_{312,1}^{[-]}\star\widehat{L}_{a,1}^{(3)}+K_{312,12}^{[+]}\star\widehat{L}_{a,12}^{(3)}+2K_{312,2}\star\widehat{L}_{a+1,2}^{(3)},\quad a=0,1,
    \end{aligned}
\end{equation}
where $\widehat{L}$ are defined by
\begin{equation}
    \begin{aligned}
        \widehat{L}_{a,1}^{(3)}=&\log\Big[\big(1+e^{-2\pi im}\widehat{Y}_{a,1}^{(3)}\big)\big(1+e^{2\pi im}\widehat{Y}_{a,1}^{(3)}\big)\Big],\\
        \widehat{L}_{a,12}^{(3)}=&\log\Big[\big(1+e^{-2\pi im}\widehat{Y}_{a,12}^{(3)}\big)\big(1+e^{2\pi im}\widehat{Y}_{a,12}^{(3)}\big)\Big],\\
        \widehat{L}_{a+1,2}^{(3)}=&\log\big(1+\widehat{Y}_{a+1,2}^{(3)}\big),\quad\widehat{L}_{a,312}^{(3)}=\log\big(1+\widehat{Y}_{a,312}^{(3)}\big).
        \end{aligned}
\end{equation}
These TBA systems are the same as those appearing in \cite[eq. (3.26)]{Ito:2024wxw} for the $(A_1,D_3)$-type AD theory, by identifying $2m-1=\ell$.

The effective central charge of the TBA system decomposes to 
\begin{equation}
    c_{{\rm eff}}=\widehat{c}_{{\rm eff},0}^{(3)}+\widehat{c}_{{\rm eff},1}^{(3)},
\end{equation}
where $\widehat{c}_{{\rm eff},a}^{(3)}$ ($a=0,1$) is defined by
\begin{equation}
    \begin{aligned}
        \widehat{c}_{{\rm eff},a}^{(3)}=&\frac{12}{\pi^{2}}\int_{-\infty}^{\infty}d\theta e^{\theta}\Big\{|m_{1}|\widehat{L}_{a,1}^{(3)}+|m_{2}|\widehat{L}_{a+1,2}^{(3)}+|m_{12}|\widehat{L}_{a,12}^{(3)}+|m_{312}|\widehat{L}_{a,312}^{(3)}\Big\}.
    \end{aligned}
\end{equation}

The Y-functions admit large $\theta$ expansions. For example,  the expansions of Y-functions corresponding to $\gamma_1$ and $\gamma_{2}$ are expressed as
\begin{equation}\label{eq:Y-max-asymptotic}
\begin{split}
-\log\widehat{Y}_{a,1}^{(3)}&\sim\left\lvert m_{1}\right\rvert e^\theta+\sum_{n=0}^\infty \widehat{m}_{a,1}^{(3,n)}e^{(1-2n)\theta},\\
-\log\widehat{Y}_{a+1,2}^{(3)}&\sim\left\lvert m_{2}\right\rvert e^\theta+\sum_{n=0}^\infty \widehat{m}_{a+1,2}^{(3,n)}e^{(1-2n)\theta},
\end{split}
\end{equation}
where $\widehat{m}_{a,s}^{(3,n)}$ is defined by
\begin{equation}
    \begin{aligned}
        \widehat{m}_{a,1}^{(3,n)}=&\frac{(-1)^{n}}{\pi}\int_{-\infty}^{\infty}d\theta e^{(2n-1)(\theta+i\phi_1)} \left(e^{i(1-2n)\phi_2} \widehat{L}_{a+1,2}^{(3)}
        +e^{i(1-2n)(\pi/2+\phi_{12})} \widehat{L}_{a,12}^{(3)}+e^{i(1-2n)(\pi/2+\phi_{312})} \widehat{L}_{a,312}^{(3)}\right),\\
        \widehat{m}_{a+1,2}^{(3,n)}=&\frac{(-1)^{n}}{\pi}\int_{-\infty}^{\infty}d\theta e^{(2n-1)(\theta+i\phi_2)}\left(e^{i(1-2n)\phi_1}\widehat{L}_{a,1}^{(3)}+e^{i(1-2n)\phi_{12}}\widehat{L}_{a,12}^{(3)}+2e^{i(1-2n)\phi_{312}}\widehat{L}_{a,312}^{(3)}\right).
    \end{aligned}
\end{equation}
We solve the TBA equations for $a=0$ numerically and compare the expansions of the solutions with the corresponding WKB expansions, as shown in Table \ref{tab:tba_wkb-max}. We find the expansion coefficients $\widehat{m}_{a,s}^{(3,n)}$ correspond to the ones of WKB expansions for quantum periods as follows:
\begin{equation}
    \widehat{m}_{0,1}^{(3,n)}=e^{i(\phi_{1}+\pi/2)(2n-1)}\Pi_{\gamma_{1}}^{(2n)}, \qquad \widehat{m}_{1,2}^{(3,n)}=e^{i\pi(2n-1)}\Pi_{\gamma_2}^{(2n)}.
\end{equation}
The relation between WKB periods and solutions to the TBA equations for $a=1$ and one-cycles $\gamma_{12}$ and $\gamma_{312}$ can be verified similarly. 

\begin{table}[]
    \centering
    \begin{tabular}{|c|c|c|}\hline
    $n$&    $\widehat{m}^{(n)}_{0,1}$ & $e^{i(2n-1)(\phi_1+i\pi/2)}\Pi_{\gamma_{1}}^{(2n)}$\\\hline
   $1$ & $0.752458431634358  -0.819383317493803i $ & $0.752458431634592 - 0.819383317493108 i$ \\\hline
   $2$&$-35.0333846702916 + 30.4727241250663i $ & $
   -35.0333846703009 + 30.4727241250714 i$  \\\hline\hline
    $n$ & $\widehat{m}_{0,2}^{(n)}$ & $e^{i(2n-1)\phi_2}\Pi_{\gamma_{2}}^{(2n)}$\\\hline
     $1$ & $-1.170472199331264 $ & $-1.170472199330518$\\\hline
   $2$&$9.82891109737608$& $9.82891109737123$ \\\hline
    \end{tabular}     \caption{
 Comparison of the coefficients in the expansions based on the TBA equations and the WKB periods at $m=2/5$ with $u_1=1/4$ and $E=1/16$. The TBA equations are solved numerically by using the discretized Fourier transformation with $2^{16}$ points and the cutoff  $(-50,50)$.}
    \label{tab:tba_wkb-max}
\end{table}

\subsection{TBA at the monomial point}
At the monomial point of the potential, i.e. $W^{\prime 2}=x^4$, the Schr\"odinger equation becomes
\begin{equation}\label{eq:Sch-eq-cubic-mon}
    \Big(-\hbar^{2}\frac{d^{2}}{dx^{2}}+x^4-2E+4\hbar mx\Big)\psi(x)=0.
\end{equation}
For $m=0$, the ODE takes the same form as in \cite{Dorey:1998pt} , where the related TBA is the $A_3/\mathbb{Z}_2$-type. 
We discuss the effect of the $\hbar$ deformation of the potential. At this monomial point, one finds that the mass parameters for the TBA equations  are
\begin{equation}
m_1=\frac{4\sqrt{2}e^{-i\pi/4}\mathbb{K}(-1)}{3}, \quad m_2=-\frac{8\mathbb{K}(-1)}{3},
\end{equation}
which provide the relations of the mass parameters: 
\begin{equation}
    \begin{aligned}
        |m_{1}|=&|m_{3}|=|m_{12}|=|m_{32}|=\frac{1}{\sqrt{2}}|m_{2}|=\frac{1}{\sqrt{2}}|m_{312}|,
        \\
        \phi_{1}=&\phi_{3}=-\frac{\pi}{4},\quad\phi_{12}=\phi_{34}=\frac{\pi}{4},\quad\phi_{2}=\pi,\quad\phi_{312}=0
        .
    \end{aligned}
\end{equation}
Substituting these data into the TBA system in the maximal chamber, one finds further identifications of the Y-functions:
\begin{equation}
    \widetilde{Y}_{a,1}^{(3)}=\widetilde{Y}_{a+2,3}^{(3)}=\widetilde{Y}_{a,12}^{(3)}=\widetilde{Y}_{a+2,32}^{(3)},\quad \widetilde{Y}_{a,2}^{(3)}=\widetilde{Y}_{a+2,2}^{(3)}=\widetilde{Y}_{a+1,312}^{(3)}=\widetilde{Y}_{a+3,312}^{(3)}.
\end{equation}
Redefining the Y-functions as
\begin{equation}
    \widehat{c}_{{\rm eff},0}^{(3)}=\frac{24}{\pi^{2}}\Big({\cal L}_{1}\big(\frac{1}{1+\widehat{Y}_{1,2}^{(3)\ast-1}}\big)+{\cal L}_{e^{2\pi im}}\big(\frac{1}{e^{2\pi im}+\widehat{Y}_{0,1}^{(3)\ast-1}}\big)+{\cal L}_{e^{-2\pi im}}\big(\frac{1}{e^{-2\pi im}+\widehat{Y}_{0,1}^{(3)\ast-1}}\big)\Big)=4(1-8m^{2}).
\end{equation}
the TBA system \eqref{eq:TBA-cubic-max} reduces to 
\begin{equation}\label{eq:TBA-D3-mon}
    \begin{aligned}
       \log\widehat{Y}_{a,1}^{(3)}=&-|m_{1}|e^{\theta}-K\star\widehat{L}_{a,1}^{(3)}-\widehat{K}\star\widehat{L}_{a+1,2}^{(3)},\\\log\widehat{Y}_{a+1,2}^{(3)}=&-|m_{2}|e^{\theta}-\widehat{K}\star\widehat{L}_{a,1}^{(3)}-2K\star\widehat{L}_{a+1,2}^{(3)},
    \end{aligned}
\end{equation}
where the kernels $K$ and $\widehat{K}$ are given by
\begin{equation}
   K=\frac{1}{2\pi}\frac{1}{\cosh(\theta)},\quad  \widehat{K}=\frac{\sqrt{2}}{\pi}\frac{\cosh(\theta)}{\cosh\big(2\theta\big)}.
\end{equation}
The equations for $a=0,1$ and $a=2,3$ form an independent set of equations. Each set forms the same $D_3$-type TBA system \cite{Zamolodchikov:1991et}. 
When $m=0$, the system is reduced to the $A_3/{\mathbb Z}_2$-TBA as in \cite{Dorey:1998pt}.

By Fourier transformation, the TBA system \eqref{eq:TBA-D3-mon}can be converted to a Y-system as follows:
\begin{equation}
    \begin{aligned}
       &\widehat{Y}_{a,1}^{(3)}(\theta-\frac{\pi i}{4})\widehat{Y}_{a,1}^{(3)}(\theta+\frac{\pi i}{4})=\frac{1}{1+\widehat{Y}_{a+1,2}^{(3)}(\theta)^{-1}},\\
       &\widehat{Y}_{a+1,2}^{(3)}(\theta+\frac{\pi i}{4})\widehat{Y}_{a+1,2}^{(3)}(\theta-\frac{\pi i}{4})=\frac{1}{\Big(1+e^{2\pi im}\widehat{Y}_{a,1}^{(3)}(\theta)^{-1}\Big)\Big(1+e^{-2\pi im}\widehat{Y}_{a,1}^{(3)}(\theta)^{-1}\Big)},
    \end{aligned}
\end{equation}
The effective central charge decomposes to 
\begin{equation}
    c_{{\rm eff}}=\widehat{c}_{{\rm eff},0}^{(3)}+\widehat{c}_{{\rm eff},1}^{(3)},
\end{equation}
where $\widehat{c}_{{\rm eff},a}^{(3)}$ is now given by
\begin{equation}\label{eq:ceff-mon}
   \widehat{c}_{{\rm eff},a}^{(3)}=\frac{24}{\pi^{2}}\int d\theta e^{\theta}\Big(|m_{1}|\widehat{L}_{a,1}^{(3)}+|m_{2}|\widehat{L}_{a+1,2}^{(3)}\Big).
\end{equation}
Using the asymptotic values of the Y-functions at $\theta\to-\infty$:
\begin{equation}
    \widehat{Y}_{0,1}^{(3)*}=\frac{1}{2\cos \frac{2\pi m}{3}}, \qquad \widehat{Y}_{1,2}^{(3)*}=\frac{\sin(\frac{2\pi m}{3})}{\sin(2\pi m)}, 
\end{equation}
the effective central charge $\widehat{c}_{{\rm eff},0}^{(3)}$ can be evaluated analytically as
\begin{equation}
     \widehat{c}_{{\rm eff},0}^{(3)}=\frac{24}{\pi^{2}}\Big({\cal L}_{1}\big(\frac{1}{1+\widehat{Y}_{1,2}^{(3)\ast-1}}\big)+{\cal L}_{e^{2\pi im}}\big(\frac{1}{e^{2\pi im}+\widehat{Y}_{0,1}^{(3)\ast-1}}\big)+{\cal L}_{e^{-2\pi im}}\big(\frac{1}{e^{-2\pi im}+\widehat{Y}_{0,1}^{(3)\ast-1}}\big)\Big)=4(1-8m^{2}).
\end{equation}
Similarly, $\widehat{c}_{{\rm eff},1}^{(3)}$ is given by
\begin{equation}
     \widehat{c}_{{\rm eff},1}^{(3)}=4(1+8m^2).
\end{equation}
The total effective central charge associated with two sets of TBA systems is
\begin{equation}
    c_{\rm{eff}}=8.
\end{equation}
This effective central charge is observed to be the same as that in the minimal chamber \eqref{eq:total-ceff}. 

\begin{figure}[h]
\begin{center}
\begin{minipage}[b]{0.33\textwidth}
\resizebox{60mm}{!}{\includegraphics{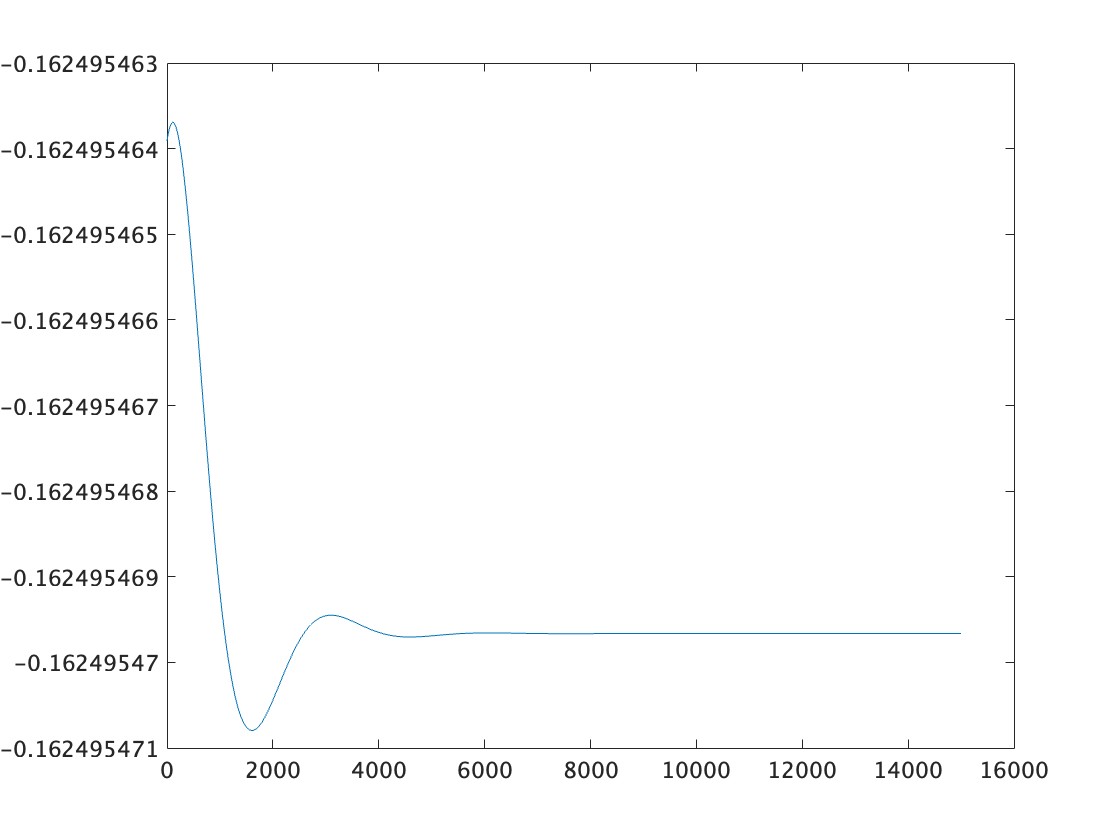}}
\subcaption{$m=0.47$}
\end{minipage}
\hspace{0.04\textwidth}
\begin{minipage}[b]{0.33\textwidth}
\resizebox{60mm}{!}{\includegraphics{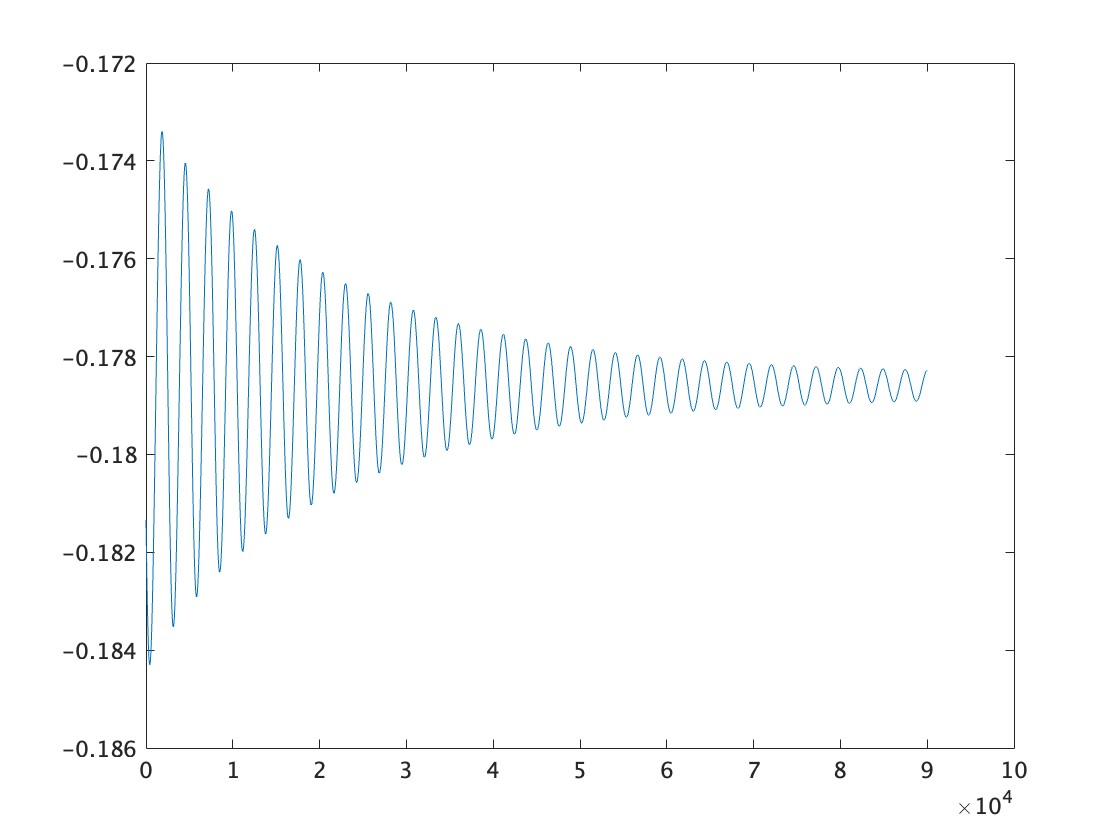}}
\subcaption{$m=0.48$}
\end{minipage}
\begin{minipage}[b]{0.33\textwidth}
\resizebox{60mm}{!}{\includegraphics{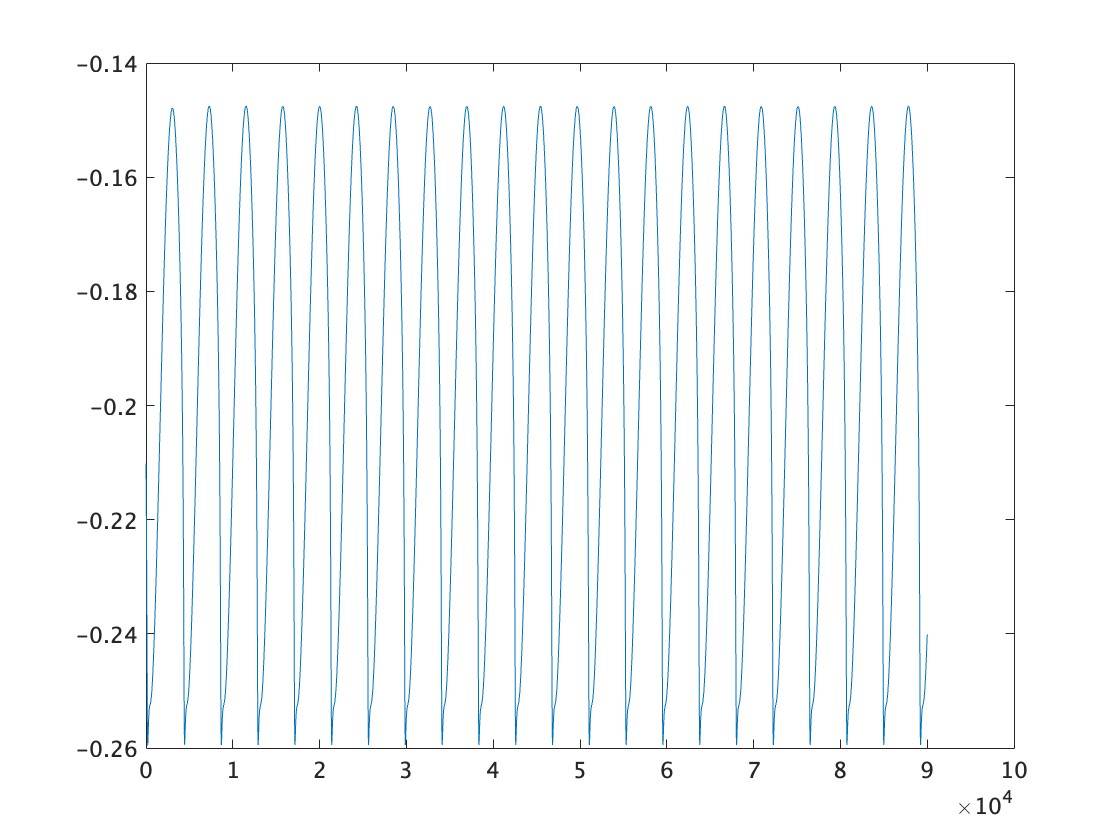}}
\subcaption{$m=0.49$}
\end{minipage}
\end{center}
\caption{ Plots of ${\rm Re}(\widehat{m}_{a,1}^{(3,1)}e^{i\pi/4})$ for $m=0.47, 0.48, 0.49$ as the function of the iteration number  $n_{\rm iter}$. The horizontal line represents $n_{\rm iter}-n_{\rm iter}^0/2$ with $n_{\rm iter}^0=10^5$ ($m=0.48, 0.49$) or $n^0_{\rm iter}=3\times 10^4$ ($m=0.47$). The TBA equations are solved numerically by a discretized Fourier transformation with $2^{16}$ points and the cutoff $(-40,40)$. }
\label{fig:iter1}
\end{figure}

The large $\theta$ expansions of the Y-functions take the form of \eqref{eq:Y-max-asymptotic}, where the expansion coefficients are defined by
\begin{equation}
    \begin{aligned}
        \widehat{m}_{a,1}^{(3,n)}=&\frac{(-1)^{n-1}}{\pi}\int_{-\infty}^{\infty}d\theta e^{(2n-1)\theta} \widehat{L}_{a,1}^{(3)}
        +\frac{\sqrt{2}(-1)^{\left\lfloor \frac{n-1}{2} \right\rfloor}}{\pi}\int_{-\infty}^{\infty}d\theta e^{(2n-1)\theta}\widehat{L}_{a+1,2}^{(3)},\\
        \widehat{m}_{a+1,2}^{(3,n)}=&\frac{2(-1)^{n-1}}{\pi}\int_{-\infty}^{\infty}d\theta e^{(2n-1)\theta}\widehat{L}_{a+1,2}^{(3)}+\frac{\sqrt{2}(-1)^{\left\lfloor \frac{n-1}{2} \right\rfloor}}{\pi}\int_{-\infty}^{\infty}d\theta e^{(2n-1)\theta}\widehat{L}_{a,1}^{(3)}.
    \end{aligned}
\end{equation}
It is worth noting that $\widehat{m}_{a,1}^{(3,n)}$ and $\widehat{m}_{a+1,2}^{(3,n)}$ are not independent:
\begin{equation}
    \sqrt{2}(-1)^{\left\lceil \frac{n-1}{2}\right\rceil }\widehat{m}_{a,1}^{(3,n)}=\widehat{m}_{a+1,2}^{(3,n)}.
\end{equation}
These expansions can be identified with the expansions of quantum periods \eqref{eq:qp1} by
\begin{equation}
    \widehat{m}_{0,1}^{(3,n)}=e^{i\pi(2n-1)/4} \Pi_{\gamma_1}^{(2n)}, \qquad \widehat{m}_{1,2}^{(3,n)}=e^{i\pi(2n-1)}\Pi_{\gamma_2}^{(2n)}.
    \label{eq:tba_wkb}
\end{equation}
Eqs. \eqref{eq:tba_wkb} can be checked numerically in the region $0\leq m\leq 0.5$.
We can see a good numerical correspondence for $0\leq m< 0.48$. The result $m=2/5$ is shown in
Table \ref{tab:tba_wkb-mono}.
In the region $0.48<m<0.5$, the numerical solution of the TBA equations shows unstable behavior, where the numerical solution does not converge to the fixed solution when the iteration number becomes large. 
However, it is interesting to see the behavior of $\widehat{m}_{a,1}^{(3,n)}$ as a function of the iteration number $n_{\rm iter}$ of the TBA equation.
For $m=0.49$, the value of ${\rm Re}(\widehat{m}_{0,1}^{(3,1)}e^{i\pi/4})$ shows the oscillating behavior with respect to the iteration number (Fig. \ref{fig:iter1}).
If one averages over a single period, one finds that the value of ${\rm Re}( \widehat{m}_{0,1}^{(3,1)}e^{i\pi/4})$ is $-0.137595790212676 $, which is close to the WKB result $-0.1379059410180667$.
The oscillation indicates that the complex poles and zeros of the Y-function approach to the integration contour, which occurs in the region $m>{1\over2}$. In this case, we need to study the analytic continuation of the TBA equations \cite{Dorey:1996re,Klummpe:1991vs,Bazhanov:1996aq,Fendley:1997ys,Gabai:2021dfi,Ito:2023cyz,Ito:2024nlt}
to investigate this issue, which is left for a future problem.

Using $\widehat{m}_{a,s}^{(3,1)}$ and ${\Pi}_\gamma^{(2)}$, the effective central charge \eqref{eq:ceff-mon} can be expressed as
\begin{equation}
    \begin{aligned}
        \widehat{c}_{{\rm eff},a}^{(3)}=&\frac{24}{\pi}|m_{1}|\widehat{m}_{a,1}^{(3,1)}=\frac{12}{\pi\sqrt{2}}(|m_{2}|\widehat{m}_{a,1}^{(3,1)}+|m_{1}|\widehat{m}_{a+1,2}^{(3,1)})=\frac{12}{\pi\sqrt{2}}(e^{\pi i/4}\Pi_{\gamma_2}^{(0) }\Pi_{\gamma_1}^{(2)}+e^{\pi i/2}\Pi_{\gamma_1}^{(0) }\Pi_{\gamma_2}^{(2)}),
    \end{aligned}
\end{equation}
which also imposes constraints on the classical and next order correction of WKB periods.
In conclusion, the monomial potential point in the maximal chamber is described by the two sets of $D_3$-type TBA equations, which are regarded as a $\mathbb{Z}_4$ extension of $A_3/\mathbb{Z}_2$ TBA equations \cite{Dorey:1998pt}.

\begin{table}[]
    \centering
    \begin{tabular}{|c|c|c|c|c|}\hline
    $n$&    $\widehat{m}^{(3,n)}_{0,1}$ & $e^{i\pi(2n-1)/4} \Pi_{\gamma_1}^{(2n)}$
    & $\widehat{m}_{1,2}^{(3,n)}$ & $ e^{i\pi(2n-1)}\Pi_{\gamma_2}^{(2n)}$\\\hline
   $1$ & $-0.059304915935657$ & $-0.059304915935579$ & $-0.083869816431602$ & $-0.083869816431492$\\\hline
   $2$&$0.092742360420868$ & $0.092742360420846$ &$-0.131157503913685$& $-0.131157503913654$ \\\hline
    \end{tabular}    \caption{
 Comparison of the coefficients in the expansions based on the TBA equations and the WKB periods at $m=2/5$ with $u_1=0$ and $E=1/2$. The TBA equations are solved numerically by using the discretized Fourier transformation with $2^{16}$ points and the cutoff  $(-50,50)$.}
    \label{tab:tba_wkb-mono}
\end{table}

\section{Conclusion and discussion}
\label{sec:con}
In this paper, we have studied the exact WKB periods in deformed supersymmetric quantum mechanics.
The exact WKB periods for any potential are described by the ${\mathbb Z}_4$-extended TBA equations, for which we have presented the general procedure of wall crossing. In the case of cubic superpotential, the Schr\"odinger equation features a double-well potential with an additional $\hbar$ correction. When all turning points are real and distinct, the $\mathbb{Z}_4$ TBA equations decouple into two independent $D_3$-type TBA equations\cite{Ito:2024nlt}. These coincide with the GMN TBA equations of $(A_1, D_3)$-type AD theory in the minimal chamber. After wall crossing, the TBA equations reduce to two independent TBA systems as well, each matching the form of the GMN TBA equations of $(A_1, D_3)$-type AD theory in the maximal chamber. At the monomial point, where $W^\prime(x)^2=x^4$, the Schr\"odinger equation takes the same form as the monomial quartic potential but with an additional $\hbar$ correction in the potential. The TBA equations for the monomial point also separate into two independent $D_3$-type TBA systems \cite{Zamolodchikov:1991et}, which provide a $\mathbb{Z}_4$ extension of the $A_3/\mathbb{Z}_2$ TBA equations. 

In the present work, we have considered the deformation parameter $m$ in the region
$|m|\leq {1\over2}$. Around $m=\pm1/2$, the poles at $1+Y=0$ start to contribute, which makes the TBA equations unstable. It is important to consider the analytic continuation of TBA equations to cross $m=\pm 1/2$, which is related to the excited state in the integrable model. See \cite{Ito:2024nlt} for the case of the minimal chamber. In our paper, we have mainly focused on the cubic superpotential. It is also interesting to consider the TBA in the maximal chamber for a more general superpotential. Imposing the symmetry at the monomial point, it is supposed to obtain a deformed TBA system of $A_{2N-1}/\mathbb{Z}_2$-type. Our results indicate a profound connection between the deformed supersymmetric quantum mechanics with cubic superpotential and the $(A_1,D_3)$-type AD theories, whose interpretation from the SW theory is an interesting open question. Moreover, the physical interpretation of the $\mathbb{Z}_4$ extension in the TBA equations within the framework of the integrable model would be interesting. The integrable model is characterized by the TBA equations. Two distinct $D_3$ TBA equations---given in \eqref{eq:TBA-min} and \eqref{eq:TBA-D3-mon}--- are connected through the wall crossing process. It would be valuable to explore how the corresponding integrable models are related, which is supposed to be a new deformation in the integrable model. It is also worth noting that the effective central charge is invariant in the process of wall crossing. In the case of massless TBA in the minimal chamber \cite{Ito:2024nlt}, the effective central charge is found to relate to the supersymmetric index in the two-dimensional ${\cal N}=2$ theories \cite{Fendley:1997ys}. It would be interesting to see the interpretation of the effective central charge from the ${\cal N}=2$ theories or the framework of the wall crossing in the gauge theory.

\subsection*{Acknowledgements}
We would like to thank Jie Gu, Yong Li,  Roberto Tateo and Hao Zou for useful discussions. 
K.I. is supported in part by Grant-in-Aid for Scientific Research 21K03570 from Japan Society for the Promotion of Science (JSPS).
H.S. is supported by the National Natural Science Foundation of China (Grant No.12405087), Henan Postdoc Foundation (Grant No.22120055) and the Startup Funding of Zhengzhou University (Grant No.121-35220049, 121-35220581). J.Y. is supported by JST SPRING, Japan Grant Number JPMJSP2106 and JPMJSP2180, and the National Natural Science Foundation of China (Grant No.12247103).

\appendix

\section{$\mathbb{Z}_4$-extended TBA equations in the maximal chamber}\label{sec:appA}
In this appendix, we present the $\mathbb{Z}_4$-extended TBA equations for the cubic superpotential. In the maximal chamber Fig. \ref{fig:wc-path}, the TBA equations become
\begin{equation}\label{eq:TBA-max-gen}
    \begin{aligned}
        \log\widetilde{Y}_{a,1}^{(3)}=&-|m_{1}|e^{\theta}+m_{a,1}^{(\frac{1}{2})}+K_{+;1,2}\star\widetilde{L}_{a+1,2}^{(3)}+K_{-;1,2}\star\widetilde{L}_{a+3,2}^{(3)}+K_{+;1,12}^{[+]}\star\widetilde{L}_{a+2,12}^{(3)}\\&+K_{-;1,12}^{[+]}\star\widetilde{L}_{a,12}^{(3)}+K_{+;1,32}^{[+]}\star\widetilde{L}_{a+2,32}^{(3)}+K_{-;1,32}^{[+]}\star\widetilde{L}_{a,32}^{(3)}
        +K_{+;1,312}^{[+]}\star\widetilde{L}_{a+2,312}^{(3)}+K_{-;1,312}^{[+]}\star\widetilde{L}_{a,312}^{(3)},\\
        \log\widetilde{Y}_{a,2}^{(3)}=&-|m_{2}|e^{\theta}+m_{a,2}^{(\frac{1}{2})}+K_{+;2,1}\star\widetilde{L}_{a+1,1}^{(3)}+K_{-;2,1}\star\widetilde{L}_{a+3,1}^{(3)}+K_{+;2,3}\star\widetilde{L}_{a+1,3}^{(3)}+K_{-;2,3}\star\widetilde{L}_{a+3,3}^{(3)}\\&+K_{+;2,12}\star\widetilde{L}_{a+1,12}^{(3)}+K_{-;2,12}\star\widetilde{L}_{a+3,12}^{(3)}+K_{+;2,32}\star\widetilde{L}_{a+1,32}^{(3)}+K_{-;2,32}\star\widetilde{L}_{a+3,32}^{(3)}\\
        &+K_{+;2,312}\star\widetilde{L}_{a+1,312}^{(3)}+K_{-;2,312}\star\widetilde{L}_{a+3,312}^{(3)}+K_{+;2,312}\star\widetilde{L}_{a+1,312}^{(3)}+K_{-;2,312}\star\widetilde{L}_{a+3,312}^{(3)},\\
        \log\widetilde{Y}_{a,3}^{(3)}=&-|m_{3}|e^{\theta}+m_{a,3}^{(\frac{1}{2})}+K_{+;3,2}\star\widetilde{L}_{a+1,2}^{(3)}+K_{-;3,2}\star\widetilde{L}_{a+3,2}^{(3)}+K_{+;3,12}^{[+]}\star\widetilde{L}_{a+2,12}^{(3)}+K_{-;3,12}^{[+]}\star\widetilde{L}_{a,12}^{(3)}\\&+K_{+;3,32}^{[+]}\star\widetilde{L}_{a+2,32}^{(3)}+K_{-;3,32}^{[+]}\star\widetilde{L}_{a,32}^{(3)}+K_{+;3,312}^{[+]}\star\widetilde{L}_{a+2,312}^{(3)}+K_{-;3,12}^{[+]}\star\widetilde{L}_{a,312}^{(3)},\\
        \log\widetilde{Y}_{a,12}^{(3)}=&-|m_{12}|e^{\theta}+m_{a,1}^{(\frac{1}{2})}+m_{a+3,2}^{(\frac{1}{2})}+K_{+;12,1}^{[-]}\star\widetilde{L}_{a,1}^{(3)}+K_{-;12,1}^{[-]}\star\widetilde{L}_{a+2,1}^{(3)}+K_{+;12,2}\star\widetilde{L}_{a+1,2}^{(3)}\\&+K_{-;12,2}\star\widetilde{L}_{a+3,2}^{(3)}+K_{+;12,3}^{[-]}\star\widetilde{L}_{a,3}^{(3)}+K_{-;12,3}^{[-]}\star\widetilde{L}_{a+2,3}^{(3)}+K_{+;12,312}^{[-]}\star\widetilde{L}_{a,312}^{(3)}+K_{-;12,312}^{[-]}\star\widetilde{L}_{a+2,312}^{(3)},\\
        \log\widetilde{Y}_{a,32}^{(3)}=&-|m_{32}|e^{\theta}+m_{a+3,2}^{(\frac{1}{2})}+m_{a,3}^{(\frac{1}{2})}+K_{+;32,1}^{[-]}\star\widetilde{L}_{a,1}^{(3)}+K_{-;32,1}^{[-]}\star\widetilde{L}_{a+2,1}^{(3)}+K_{+;32,2}\star\widetilde{L}_{a+1,2}^{(3)}\\&+K_{-;32,2}\star\widetilde{L}_{a+3,2}^{(3)}+K_{+;32,3}^{[-]}\star\widetilde{L}_{a,3}^{(3)}+K_{-;32,3}^{[-]}\star\widetilde{L}_{a+2,3}^{(3)}
        +K_{+;32,312}^{[-]}\star\widetilde{L}_{a,312}^{(3)}+K_{-;32,312}^{[-]}\star\widetilde{L}_{a+2,312}^{(3)},\\
        \log\widetilde{Y}_{a,312}^{(3)}=&-|m_{312}|e^{\theta}+m_{a,3}^{(\frac{1}{2})}+m_{a,1}^{(\frac{1}{2})}+m_{a+3,2}^{(\frac{1}{2})}+K_{+;312,1}^{[-]}\star\widetilde{L}_{a,1}^{(3)}+K_{-;312,1}^{[-]}\star\widetilde{L}_{a+2,1}^{(3)}+K_{+;312,3}^{[-]}\star\widetilde{L}_{a,3}^{(3)}\\&+K_{-;312,3}^{[-]}\star\widetilde{L}_{a+2,3}^{(3)}+K_{+;312,2}\star\widetilde{L}_{a+1,2}^{(3)}+K_{-;312,2}\star\widetilde{L}_{a+3,2}^{(3)}+K_{+;312,2}\star\widetilde{L}_{a+1,2}^{(3)}+K_{-;312,2}\star\widetilde{L}_{a+3,2}^{(3)}\\&+K_{+;312,12}^{[+]}\star\widetilde{L}_{a+2,12}^{(3)}+K_{-;312,12}^{[+]}\star\widetilde{L}_{a,12}^{(3)}+K_{+;312,32}^{[+]}\star\widetilde{L}_{a+2,32}^{(3)}+K_{-;312,32}^{[+]}\star\widetilde{L}_{a,32}^{(3)},
    \end{aligned}
\end{equation}
where $a\equiv a+4$ and the new masses are defined by \eqref{eq:new-mass-max}. The effective central charge becomes
\begin{equation}\label{eq:ceff-max}
    c_{{\rm eff}}=\frac{6}{\pi^{2}}\int_{-\infty}^{\infty}d\theta e^{\theta}\sum_{a=0}^{3}\Big\{\sum_{s=1}^{3}|m_{s}|\widetilde{L}_{a,s}(\theta)+|m_{12}|\widetilde{L}_{a,12}(\theta)+|m_{23}|\widetilde{L}_{a,23}(\theta)+|m_{312}|\widetilde{L}_{a,312}(\theta)\Big\}.
\end{equation}
For the cubic superpotential, we find the relations \eqref{eq:mass-rel-max}, which lead to the identifications \eqref{eq:iden-max}. The TBA equations \eqref{eq:TBA-max-gen} thus reduce to two independent TBA systems
\begin{equation}\label{eq:TBA-max-dou-1}
    \begin{aligned}
        \log\widetilde{Y}_{a,1}^{(3)}=&-|m_{1}|e^{\theta}+m_{a,1}^{(\frac{1}{2})}+K_{1,2}\star\widetilde{L}_{a+1,2}^{(3)}+K_{1,312}^{[+]}\star\widetilde{L}_{a,312}^{(3)}
        +K_{1,12}^{[+]}\star\widetilde{L}_{a+2,12}^{(3)}+K_{1,12}^{[+]}\star\widetilde{L}_{a,12}^{(3)},\\
        \log\widetilde{Y}_{a+2,1}^{(3)}=&-|m_{1}|e^{\theta}-m_{a,1}^{(\frac{1}{2})}+K_{1,2}\star\widetilde{L}_{a+1,2}^{(3)}+K_{1,312}^{[+]}\star\widetilde{L}_{a,312}^{(3)}
        +K_{1,12}^{[+]}\star\widetilde{L}_{a,12}^{(3)}+K_{1,12}^{[+]}\star\widetilde{L}_{a+2,12}^{(3)},\\
        \log\widetilde{Y}_{a,12}^{(3)}=&-|m_{12}|e^{\theta}+m_{a,1}^{(\frac{1}{2})}+K_{12,1}^{[-]}\star\widetilde{L}_{a,1}^{(3)}+K_{12,1}^{[-]}\star\widetilde{L}_{a+2,1}^{(3)}
        +K_{12,2}\star\widetilde{L}_{a+1,2}^{(3)}+K_{12,312}^{[-]}\star\widetilde{L}_{a,312}^{(3)},\\
        \log\widetilde{Y}_{a+2,12}^{(3)}=&-|m_{12}|e^{\theta}-m_{a,1}^{(\frac{1}{2})}+K_{12,1}^{[-]}\star\widetilde{L}_{a+2,1}^{(3)}+K_{12,1}^{[-]}\star\widetilde{L}_{a,1}^{(3)}
        +K_{12,2}\star\widetilde{L}_{a+1,2}^{(3)}+K_{12,312}^{[-]}\star\widetilde{L}_{a,312}^{(3)},\\
        \log\widetilde{Y}_{a+1,2}^{(3)}=&-|m_{2}|e^{\theta}+K_{2,1}\star\widetilde{L}_{a+2,1}^{(3)}+K_{2,1}\star\widetilde{L}_{a,1}^{(3)}
        +K_{2,12}\star\widetilde{L}_{a+2,12}^{(3)}+K_{2,12}\star\widetilde{L}_{a,12}^{(3)}+2K_{2,312}\star\widetilde{L}_{a,312}^{(3)},\\
        \log\widetilde{Y}_{a,312}^{(3)}=&-|m_{312}|e^{\theta}+K_{312,1}^{[-]}\star\widetilde{L}_{a,1}^{(3)}+K_{312,1}^{[-]}\star\widetilde{L}_{a+2,1}^{(3)}\\&\qquad\qquad\quad
        +K_{312,12}^{[+]}\star\widetilde{L}_{a+2,12}^{(3)}+K_{312,12}^{[+]}\star\widetilde{L}_{a,12}^{(3)}+2K_{312,2}\star\widetilde{L}_{a+1,2}^{(3)},\quad a=0,1.
    \end{aligned}
\end{equation}

\bibliographystyle{ytphys.bst}
\bibliography{refs}

\end{document}